\documentclass[journal,10pt,]{IEEEtran}
\usepackage[T1]{fontenc}

\usepackage{cite}
\ifCLASSINFOpdf
\else
\fi
\usepackage{amsmath}
\usepackage[pdftex]{graphicx}
\graphicspath{ {./Images/} }
\usepackage[cmintegrals]{newtxmath}
\usepackage{gensymb}
\begin{document}
\title{Early Warning of mmWave Signal Blockage and AoA Transition Using sub-6 GHz Observations}

\author{Ziad~Ali,~\IEEEmembership{Student Member,~IEEE,} Alexandra Duel-Hallen,~\IEEEmembership{Fellow,~IEEE,} and~Hans~Hallen,~\IEEEmembership{Fellow,~APS}
\thanks{Z. Ali and A. Duel-Hallen are with the Department of Electrical and Computer Engineering, North Carolina State University, Raleigh,
NC, 27606.}
\thanks{H. Hallen is with the Physics Dept., NC State Univ., Raleigh, NC, 27695.}}

\markboth{IEEE Communications Letters~2019,~DOI:~10.1109/LCOMM.2019.2952602}{}
%

\maketitle

\begin{abstract}
The susceptibility of millimeter-wave (mmWave) signals to physical blockage and abrupt signal strength variations presents a challenge to reliable 5G communication. This work proposes and examines the feasibility of utilizing lower-frequency signals as early-warning indicators of mobile mmWave signal blockage or recovery. A physics-based channel simulation tool incorporating Fresnel diffraction and image sources is employed to demonstrate that sub-6 GHz signals "lead" mmWave signals in reaching a specific signal-strength threshold by several to tens of milliseconds at mobile speeds, suggesting early-warning systems are viable. This predictive approach stems from frequency-dependent properties of diffraction and does not assume a specific topology or mobile and obstacle speeds. Realistic simulations that include transitions from line of sight (LoS) to non-line of sight (NLoS) and reflection scenarios are employed to verify the proposed prediction capabilities. Moreover, prediction of the strongest multipath component and its angle of arrival (AoA) using sub-6 GHz observations is investigated.
\end{abstract}

\begin{IEEEkeywords}
mmWave, blockage prediction, Fresnel diffraction, physical channel model, shadowing, hybrid communications
\end{IEEEkeywords}

\IEEEpeerreviewmaketitle

\section{Introduction}
\IEEEPARstart{T}{he} emerging generation of cellular communications\footnote{This paper has supplementary "media" material provided by the authors. This includes a 2 MB pdf file with further explanations and derivations, cited as [SM \S section]. The model code is at a link from www.physics.ncsu.edu/optics}, 5G, will utilize frequencies in the range of 30 - 300 GHz (mmWave) to overcome the bandwidth limitations inherent to current 4G systems (sub-6 GHz) \cite{Modeling}. However, mmWave communication signals do not diffract as strongly and thus are more susceptible to blocking by physical objects than sub-6 GHz signals
 \cite{Propagation}. 
Such effects are especially detrimental in mobile communications 
or with moving obstacles \cite{BeamSpy,MachineLearn,Tennessee}.  

In this paper, we explore the utilization of sub-6 GHz-band channel state information (CSI) to provide early warning of blockage and other received signal strength (RSS) variations of mmWave signals \em far ahead into the future\em. The proposed method is suitable for hybrid communication systems, where the sub-6 GHz and mmWave bands are employed simultaneously \cite{Steering,OutOfBand,Angular}. 
Using the fact that diffracted sub-6 GHz signals reach a specified RSS threshold much earlier than mmWave signals \cite{Optics},we employ the threshold crossing of the former to forecast blockages and other rapid changes in the latter several to tens of msec (several to hundreds of slots) ahead in mobile communications systems. The proposed predictive approach would provide hybrid mobile communication systems with sufficient time to adapt the data rate, search for a new beam, or perform a handover between the two frequencies before a significant change of the mmWave signal occurs. The insights and results in this paper are based on our physical model \cite{Prediction,Template, WirelessLRPSigProcMag2000}, which uses the method of images and Fresnel diffraction \cite{Optics}. Our analysis and simulation results demonstrate the proposed early warning capability in typical mobile communication environments. Moreover, we show that the strongest multipath component of the mmWave signal and its AoA can be predicted using sub-6 GHz observations, thus aiding a mmWave beam search.

\textbf{Related Work} Research on \em predicting blockage \em includes using beams with overlapping zones to create correlations that aid in beam selection in response to a blockage \cite{BeamSpy}. However, this method does not provide early warning of an upcoming blockage. A machine learning-based method \cite{MachineLearn} uses training data on base station handoffs in a static environment, identical to that of the mobile user, to predict that user's handoffs, presumably due to blockage of LoS. A handover method \cite{Tennessee} uses blockage of neighboring paths to predict loss of the primary one, but requires slowly moving obstacles,  a scattering-rich environment, and tracking of several mmWave antenna beam formations. In contrast to \cite{MachineLearn,Tennessee}, the method described in this paper does not assume a specific environment, obstacle motion, or speeds. 

While the references above employ only in-band mmWave channel observations, \em out-of-band \em information was exploited in \cite{Steering, OutOfBand, Angular, EstimatingChannels, AA}. However, we compare the mmWave and sub-6 GHz signals \emph{at different time instances (or locations)}, not at the same time instance as in the referenced works. These references aim to reduce the antenna beam search or detect existing blockage at mmWave, not warn of an upcoming blockage or an abrupt RSS or AoA change. Our results in section III.C corroborate the findings in \cite{OutOfBand, EstimatingChannels, AA}, but \em the main contribution of this paper is demonstration of the early warning capability of the sub-6 GHz observations, which has not been reported previously in the literature.\em  

Finally, the physical model used used in this paper was validated using measured data in \cite{Prediction, WirelessLRPSigProcMag2000}. These results pertained to long-range prediction (LRP) of multipath fading at sub-6GHz frequencies and demonstrated that the model captured the channel variation in a more realistic way than conventional, stationary channel models. In this paper, the same physical model accurately characterizes the hybrid channel and the early warning capability of sub-6 GHz signals. We emphasize that this model differs significantly from ray tracing \cite{RayTracing}, statistical \cite{Angular, EstimatingChannels}, geometric \cite{OutOfBand}, and point cloud models \cite{PointCloud}. First, one key element in mmWave propagation is the importance of small reflectors, such as sign posts, since their sizes are comparable or larger than the mmWave wavelength and thus they might produce strong reflections. A very dense set of rays must be used and placed carefully in a ray-tracing model to ensure that at least some rays intercept the small reflectors \cite{RayTrace}. Since our model is based on calculation from every reflector, a reflection is never missed and always handled accurately. Second, while the hybrid channel model in \cite{OutOfBand} relied on qualitative observations from measurements, our model accurately captures frequency-dependent diffraction even close to an object, uses full Fresnel formalism to avoid quantization onto rays, includes diffraction modification to the non-reflected component, and models small reflectors, which are not addressed in \cite{UTD} or enhanced ray tracing \cite{RayTracing}. 


\section{Physical Channel Modeling}

Consider an omnidirectional, narrowband transmission. The equivalent low-pass received signal at a frequency $f_k$ is given by $r_k(t) = h_k(t) s(t) + n(t)$, where $s(t)$ is the transmitted signal, $n(t)$ is additive noise, and
\begin{equation} \begin{split}
h_k(t) & = \sum_{m=0}^{M} h_{k,m}(t) \text{ with }\\
h_{k,m}(t) & = A_{k,m}(t)e^{j(2\pi f_{dm,k}t\, cos(AoA_m(t)) + \phi_{k,m}(t))} \label{eqnh}
\end{split} \end{equation}
is the complex channel coefficient, where $h_{k,0}(t)$, is the 'direct,' or non-reflected, multipath component, $h_{k,m}(t)$ for $m>0$ corresponds to reflections in the modeled environment from the $M$ reflectors, $A_{k,m}(t)$ and $\phi_{k,m}(t)$ are the time-dependent envelope and phase, respectively, of the $m^{th}$ multipath component, and $f_{dm,k} = f_kv/c$ is the maximum Doppler shift for a mobile with speed $v$ and speed of light $c$. The angle of arrival $AoA_m(t)$ is the angle between the reflector-to-mobile or the transmitter-to-mobile direction and mobile direction for $m>0$ and $m=0$, respectively.
The $A_{k,m}(t)$ and $\phi_{k,m}(t)$ account for the propagation delays of each reflector $m$ and phase shifts in the diffraction process. 
We employ an image and Fresnel diffraction-based calculation tool \cite{Prediction, Template} to model and simulate the frequency-dependent coefficient $h_k(t)$ in (\ref{eqnh}) in various channel environments. While this earlier work employed the model to enable long-range prediction on multipath fading \cite{WirelessLRPSigProcMag2000} and UWB template design \cite{Template}, in this paper we use the physical model to characterize the hybrid channel and to investigate frequency-dependent diffraction effects. To achieve this goal, we have updated the calculation tool to perform two diffraction calculations: one from the transmitter to the reflector, and one from the effective source to the receiver. The model naturally and accurately transforms the $m$=0 (non-reflected) component $h_{k,0}(t)$ in (\ref{eqnh}) from the LoS to a NLoS (diffracted) signal or vice-versa. Identically-sized and located apertures substitute reflectors, while effective sources are created and positioned using the image method as in \cite{Prediction}.

\begin{figure}[!t]
\centering
a.{\includegraphics[scale = 0.25]{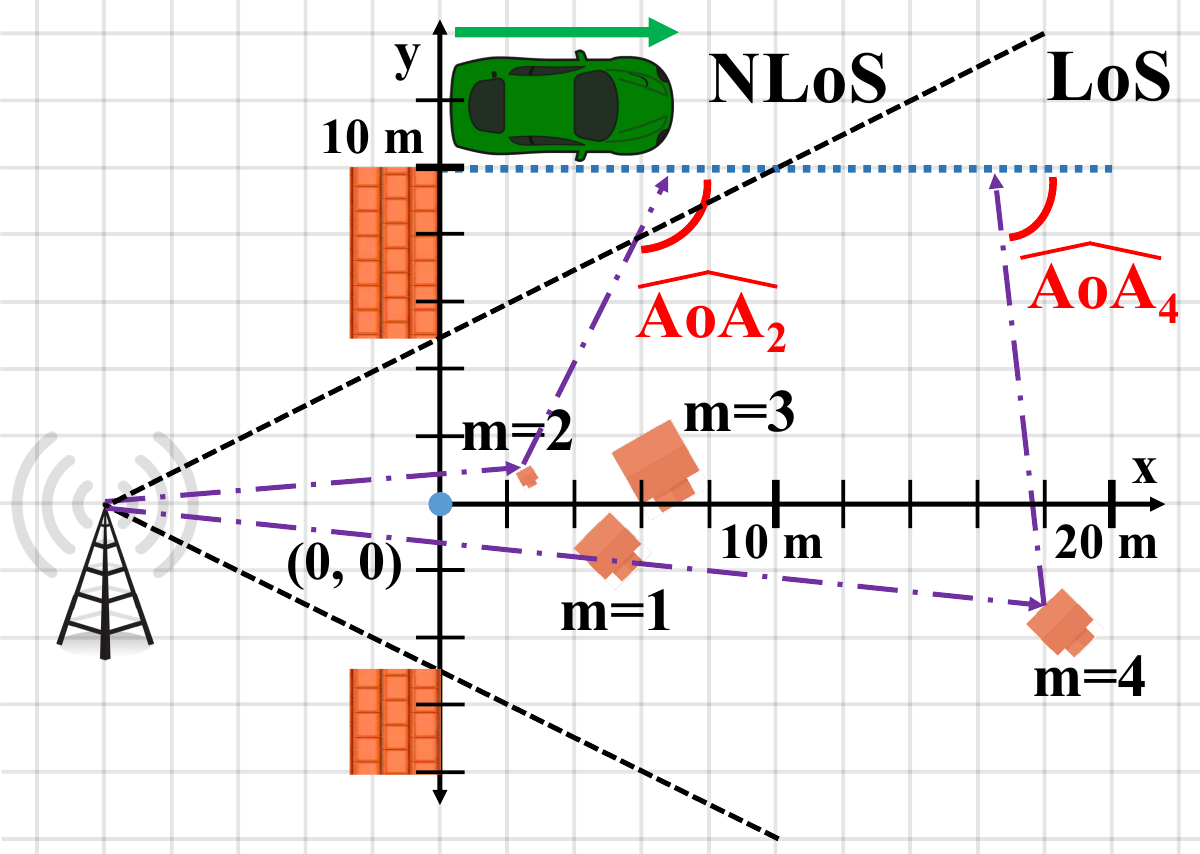}}
b.{\includegraphics[scale = 0.15]{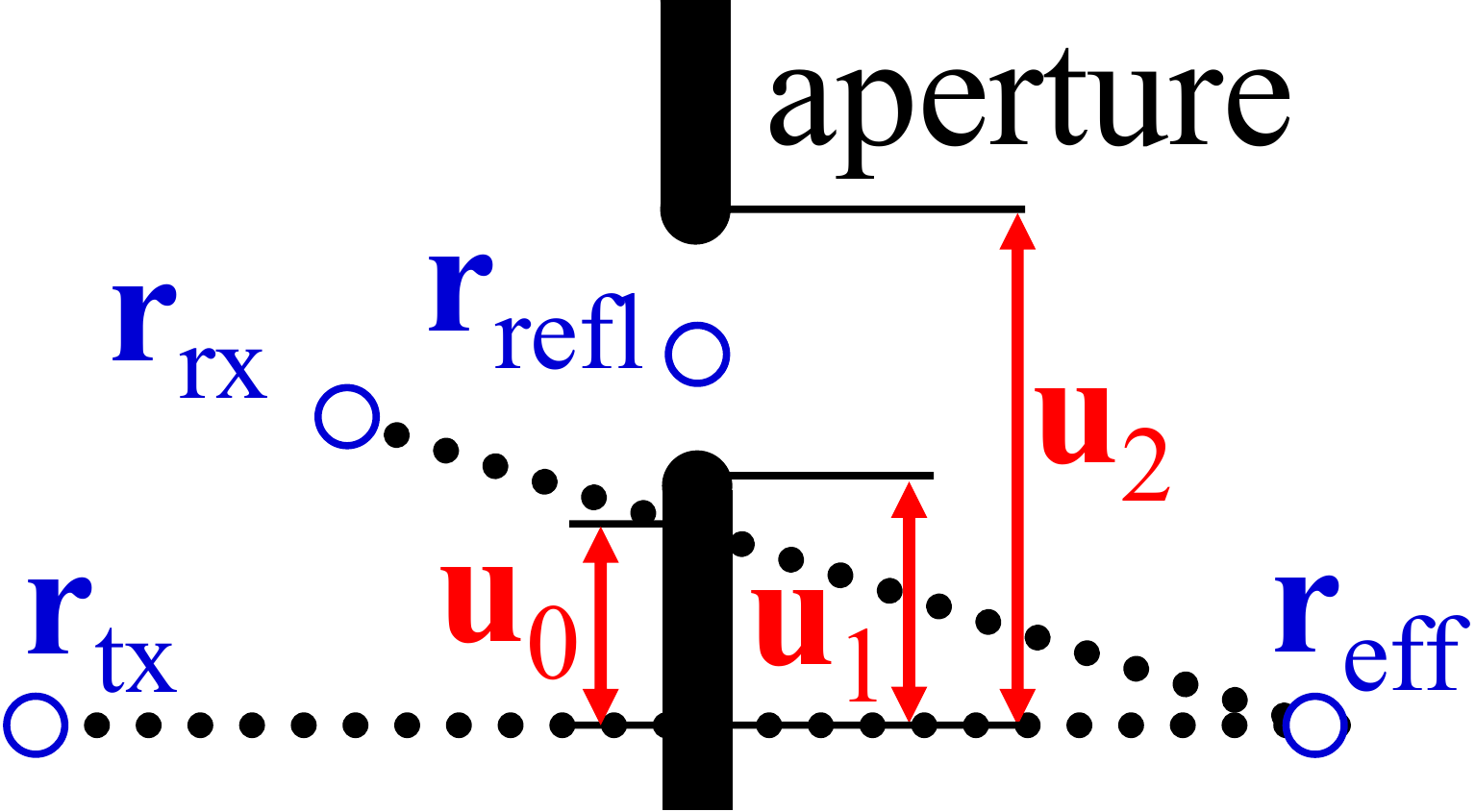}}

c.{\includegraphics[width=0.21\textwidth]{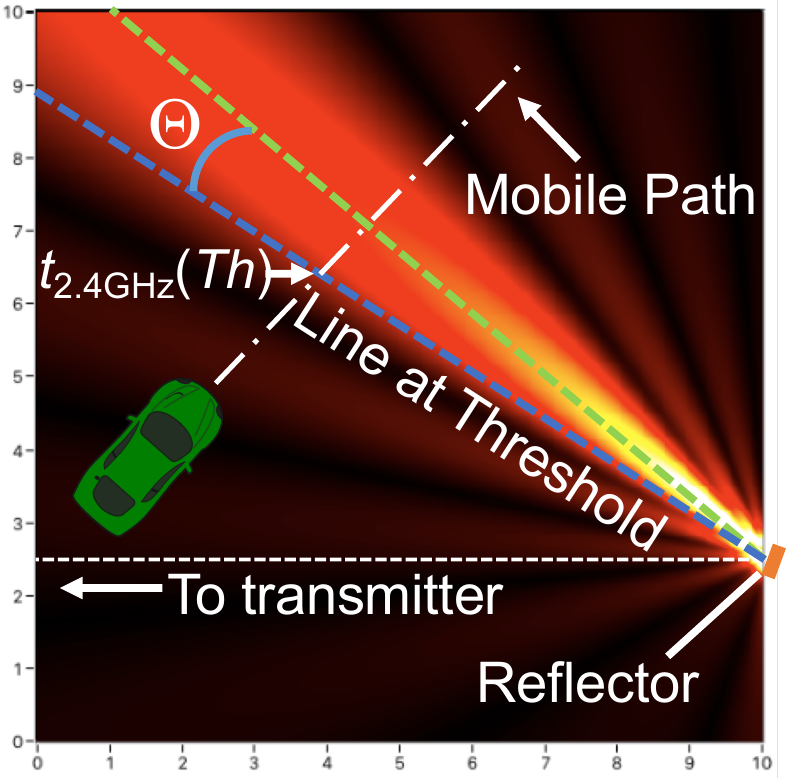}}
{\includegraphics[scale = 0.22]{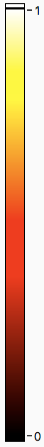}}
d.{\includegraphics[width=0.21\textwidth]{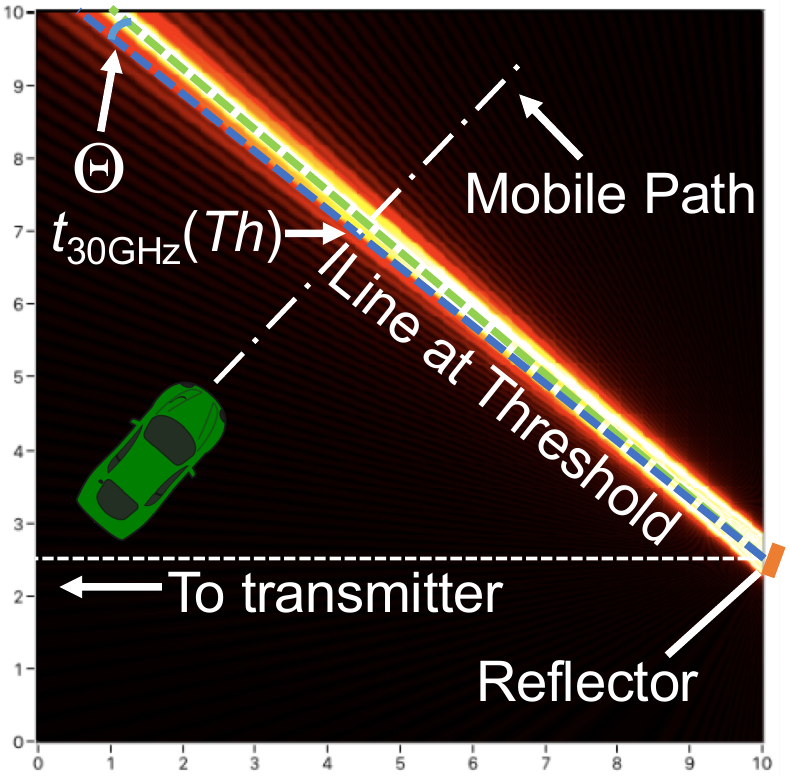}}
\hfil
\caption{Model environment, diffraction diagrams, and example spatial patterns. (a) Simulated scenarios of car movement from NLoS to LoS and reflection into NLoS region. Transmitter, its aperture (brick walls), four reflectors $m=1-4$, and the channel calculation path ($\mathbf{r}_{rx}$ values during the car's motion) are shown. (b) A diagram to define the diffraction parameters for one reflector (at aperture). (c) An example of the spatial pattern, $|h_{k,m}(x, y)|$, in a  10 m x 10 m region with a 2.4 GHz transmitter and the diffraction angle $\theta$, mobile path, \emph{line at threshold} $|h_{k,m}(x, y)|=Th$, and position $t_{2.4GHz}(Th)$ in (\ref{tauth}) indicated. (d) The same scenario as (c), but $f_k$ = 30 GHz.}
\label{LoStoNLoSDiagram}
\end{figure}

The model calculates the spatial coefficient $h_{k,m}(x, y)$ point-by-point along a specific trajectory for a chosen configuration of \textit{M} curved or flat reflectors of different sizes. It can be converted to time for straight-line mobile motion $h_{k,m}(t) = h_{k,m}(x(t), y(t)) = h_{k,m}(x_0+v_xt, y_0+v_yt)$ in (\ref{eqnh}) via the receiver's starting ($t=0$) position $(x_0,y_0)$ and speed $\textbf{v} = (v_x,v_y)$. Curved reflectors are relatively unimportant compared to LoS or even diffraction from a small, flat reflector except when the mobile is adjacent to the curved reflector [SM \S III], so they are not included in our scenarios. In addition, while the model is capable of 3-d, a 2-d simulation is used in this work since 3-d is more complex but unlikely to result in qualitative differences [SM \S III]. Scenarios with a LoS component along the entire path are not considered since the LoS component dominates, making these paths useful but not interesting. 

An example of a model scenario is shown in Fig. \ref{LoStoNLoSDiagram}(a).
The transmitter aperture is given by the gap in the brick wall.
A receiver (car) moves from NLoS to LoS along a horizontal path. Four reflectors $m=1-4$ are shown in this example ($M$=4). For each reflector $m$ in the general $M$ case, we define the parameters for the calculation, including its location and the effective source location, as in Fig. \ref{LoStoNLoSDiagram}(b). The coordinates in the (reflector or transmitter) aperture plane are denoted by $u, v$, with $v$ normal to the page, and are measured relative to the intersection of that plane with the straight line from the effective source location $\mathbf{r}_{eff,m}$ to the mobile location $\mathbf{r}_{rx}$. Each reflector $m$ at location $\mathbf{r}_{refl,m}$ has reflectance $\Re_m$ and phase $\phi_m$. All positions, $\mathbf{r}_{rx}$, $\mathbf{r}_{refl,m}$, $\mathbf{r}_{eff,m}$, $\mathbf{r}_{tx}$, $\mathbf{r}_{txaper}$, can be varied along individual straight-line paths, and reflectors rotated, although in this work we only vary $\mathbf{r}_{rx}$ within each model evaluation. The two back-back Fresnel diffractions \cite{Optics} from a point source then give the electric field complex amplitude at $\mathbf{r}_{rx}$:
\begin{equation} \begin{split}
& h_{k,m}(\mathbf{r}_{rx}) =  \frac{-0.25\Re_m e^{j(2\pi r_m f_k/c + \phi_{k,m})}|\mathbf{r}_{refl,m}-\mathbf{r}_{eff,m}|}{|\mathbf{r}_{refl,m}-\mathbf{r}_{eff,m}|+|\mathbf{r}_{rx}-\mathbf{r}_{refl,m}|}\\
& * \frac{Fr(w_{u_1}, w_{u_2}) Fr(w_{v_1}, w_{v_2}) Fr(w_{u_{1tx}}, w_{u_{2tx}}) Fr(w_{v_{1tx}}, w_{v_{2tx}})}{|\mathbf{r}_{txaper}-\mathbf{r}_{tx}|+|\mathbf{r}_{refl,m}-\mathbf{r}_{txaper}|}.   \label{eqnE}
\end{split} \end{equation}
The Fresnel Integral $Fr$, and definitions, are given by
\begin{equation} 
\begin{split} Fr(w_{u_1}, w_{u_2}) = C(w_{u_2})-C(w_{u_1})-jS(w_{u_2})+jS(w_{u_1}),\\
C(w)=\int^w_0{cos(\frac{\pi q^2}{2})dq} \text{ and } S(w)=\int^w_0{sin(\frac{\pi q^2}{2})dq},
\end{split}
\label{eqnFr}
\end{equation}
\begin{equation}
w_{\xi_\zeta,k,m} = \sqrt{\frac{2f_k}{c\rho_m}}(\xi_{\zeta,m}-\xi_{0,m}) =  \sqrt{\frac{2}{\lambda_k\rho_m}}(\xi_{\zeta,m}-\xi_{0,m}) \label{eqnw}
\end{equation}
for $\xi \in \{u, v\} \text{ and } \zeta \in \{1, 2, 1tx, 2tx\}$. All $w$ parameters in (\ref{eqnE}) and (\ref{eqnFr}) have a $k,m$ dependence as in (\ref{eqnw}), $\lambda_k$ the wavelength, and $\rho_m$ a distance parameter. For the second diffraction, 
\begin{equation}
\frac{1}{\rho_m} = \frac{1}{|\mathbf{r}_{refl,m}-\mathbf{r}_{eff,m}|}+\frac{1}{|\mathbf{r}_{rx}-\mathbf{r}_{refl,m}|}. \label{eqnrho}
\end{equation}
In 2-d, the aperture is only along the \textit{u}-direction, so the \textit{v}-direction does not have diffraction, and thus $v_1=-\infty, v_2=\infty$, which causes $C(w_{v_2})=S(w_{v_2})=-C(w_{v_1})=-S(w_{v_1})=0.5$ and $Fr(w_{v_1}, w_{v_2})=Fr(w_{v_{1tx}}, w_{v_{2tx}})=(1-j)$ for all $m$.

Next, we illustrate the frequency-dependent properties of diffracted signals and provide insight into the early warning capabilities of sub 6-GHz signals. In Fig. \ref{LoStoNLoSDiagram}(c, d), the spatial images of $|h_{k,1}(x, y)|$ are shown at sub-6 GHz and mmWave frequencies, respectively, for the same 0.5 m long reflector rotated 160$^\circ$ counter-clockwise from the x-axis and positioned 0.1 m to the right and 2.5 m above the bottom of the 10 m x 10 m diffracted-signal region shown. The specular reflection of the transmitter, which is 1000 m to the left, is along the direction of the bright line through the pattern. The color scale is 0-1 (numerical max is $\sim$1.3 in the region) referenced to the transmit signal strength. As the receiver moves along the line shown or other paths through the region, we observe an earlier and more gradual transition in RSS at lower frequencies than at higher frequencies. To quantify this observation, consider the set of points $(x, y)$ for which $|h_{k}(x, y)| = Th$, where $Th$ is the \emph{threshold}. Given the trajectory and speed, we define the \emph{threshold delay} for frequencies $f_1$ and $f_2 > f_1$ as
\begin{equation}
\tau(Th) = t_{f_2}(Th) - t_{f_1}(Th) \;\;\;\; (seconds), \label{tauth}
\end{equation}
where $t_{f_k}(Th)$ is the time when $|h_k(t)|$ = $Th$ for frequency $f_k$. The threshold delay can be expressed as a distance by multiplying the terms of (\ref{tauth}) by the mobile's speed. In this paper, we assume a vehicular speed of 20 m/s and $f_1$ = 2.4 GHz, $f_2$ = 30 GHz.
In Fig. \ref{LoStoNLoSDiagram}(c, d), only the reflected component $m=1$ is shown. The set of points where $|h_{k,1}(x, y)| = Th$ forms a \emph{line at threshold} (sufficiently far from the reflector) at an angle from specular reflection, which we call the diffraction angle: $\theta_{diffraction angle}(Th) = w(Th)\sqrt{c r_m/(2f_kr_{1,m}r_{2,m})}$ where $r_{1,m} = |\mathbf{r}_{refl,m}-\mathbf{r}_{eff,m}|$, $r_{2,m} = |\mathbf{r}_{rx}-\mathbf{r}_{refl,m}|$, and $r_m = |\mathbf{r}_{rx}-\mathbf{r}_{eff,m}|$ [SM \S IV]. The $w(Th)$, as in (\ref{eqnw}) is a frequency-independent constant determined from $Th$ via inversion of the Fresnel terms (\ref{eqnFr}) in (\ref{eqnE}). Note that the \emph{diffraction angle decreases with frequency}. Thus, the mobile path intersects the $Th$ (dotted line) earlier for 2.4 GHz ($t_{2.4 GHz}(Th)$, Fig. \ref{LoStoNLoSDiagram}(c)) than at 30 GHz ($t_{30 GHz}(Th)$,  Fig. \ref{LoStoNLoSDiagram}(d)), illustrating the early warning potential of the sub-6 GHz signals. While the above discussion focused on the reflected ($m>0$) components in (\ref{eqnh}), the LoS/NLoS transition ($m=0$) has similar dependencies.

\section{Simulation Results}
In this section, we employ our physics-based simulation tool (section II) to demonstrate the capability of lower-frequency observations to predict mmWave signal-strength changes. The RSS values $|h_k(t)|$ in (1) were calculated for $f_1 = 2.4$ GHz and $f_2 = 30$ GHz for several topologies chosen to highlight typical propagation scenarios and challenging cases for the proposed 'early warning' concept. For each topology, we computed $\tau(Th)$ in (\ref{tauth}) for appropriately chosen threshold values. We also provide a statistical analysis that demonstrates the impact of reflector roughness on early-warning capabilities.

\begin{figure}[!t]
\centering
{\includegraphics[scale = 0.32]{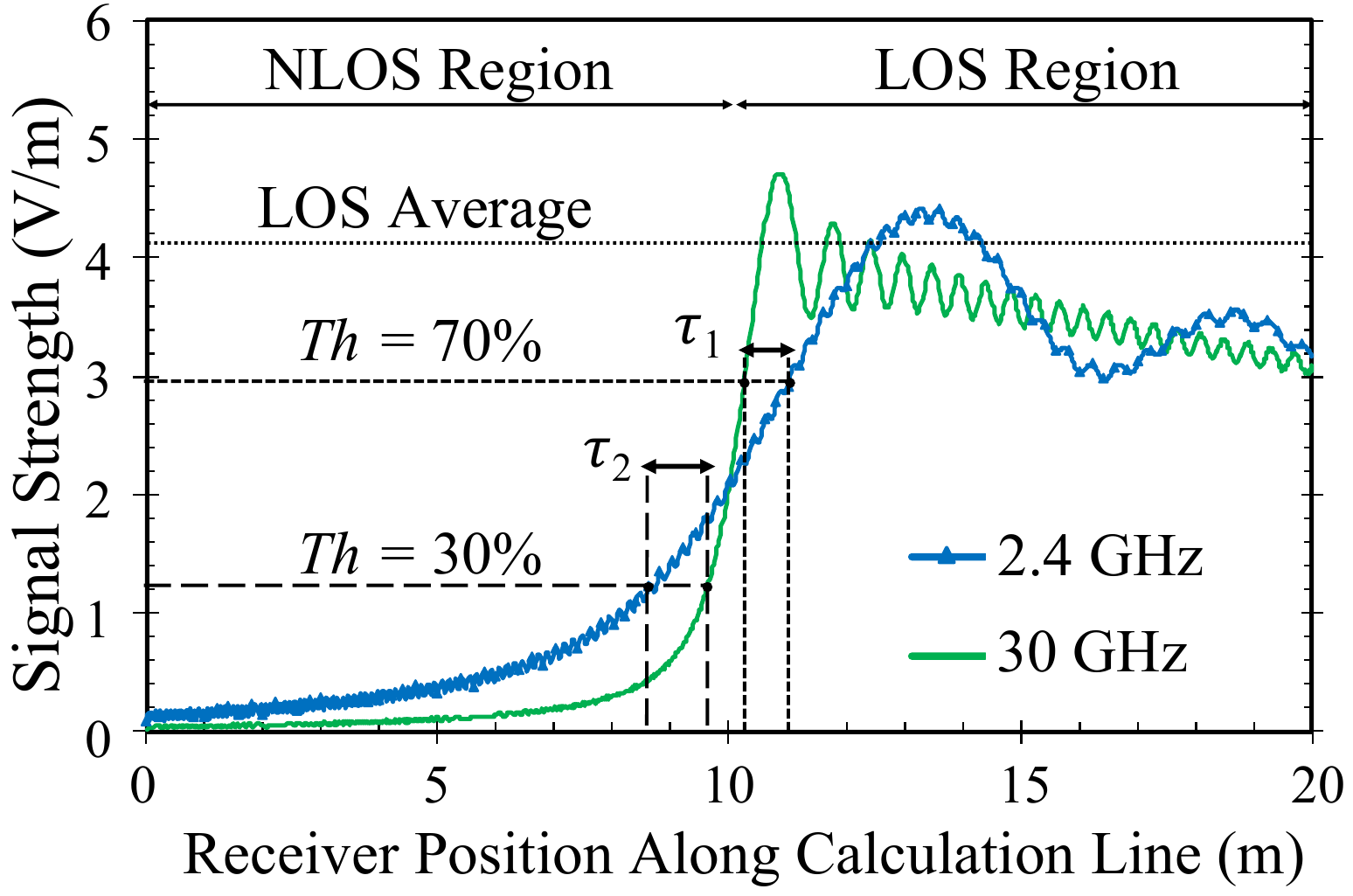}}
\hfil
\caption{RSS $|h_{k}(x, y)|$ values at $y$ = 10 m, $f_1 =$ 2.4 GHz, $f_2 =$ 30 GHz along a 20 m path from the scenario in Fig. \ref{LoStoNLoSDiagram}(a), but with reflectors removed. Delay $\tau_{1}(Th=70\% LoS)$ = 38 ms (0.76 m) (\ref{tauth}) for LoS/NLoS transition (mobile moves right to left) and $\tau_{2}(Th=30\% LoS)$ = 44 ms (0.88 m) for NLoS/LoS (left to right motion). Mobile speed = 20 m/s.}
\label{AllSignalStrengthData}
\end{figure}

\subsection{LoS to NLoS Transition: Blockage}
This section is devoted to sudden blockage of the mmWave LoS signal due to poor diffraction around obstacles in the environment, one of the primary limiting factors of base station coverage and causes of interrupted mmWave communication \cite{Modeling, Propagation}. Figure \ref{LoStoNLoSDiagram}(a) with reflectors removed ($M=0$ in (\ref{eqnh})) illustrates a scenario typical in blockage. The receiver is moving from the LoS to the NLoS region (right to left) along the horizontal dashed line. A brick building (the transmitter aperture) eventually blocks the LoS, and the received signal is determined by the diffraction. Figure \ref{AllSignalStrengthData} depicts the RSS for both frequencies along the path, and the lower frequency is impacted much earlier than the mmWave frequency as discussed in section II. We also observe the more abrupt transition to NLoS at the mmWave frequency, illustrating sudden blockage. The threshold $Th$ is chosen to be 70\% of the average LoS value for this LoS/NLoS transition, and $\tau$ = 38 ms (0.76 m) (\ref{tauth}). This threshold delay is \emph{sufficiently long to provide an early warning} (tens to hundreds of slots) of the upcoming transition to NLoS and blockage of the mmWave signal using observations at the lower band, \emph{even in vehicular communications}. Note that other threshold values can be chosen, but $Th$ = 50\% should not be used since the RSS profiles at all frequencies intersect at 50\% for edge diffraction, where the Fresnel integral argument $w$ (\ref{eqnw}) is zero [SM \S II]. Moreover, when the mobile moves from left to right in Fig. \ref{LoStoNLoSDiagram}(a), i.e., a NLoS to LoS transition (signal recovery), $Th$ = 30\% of LoS RSS is used in (\ref{tauth}), and the threshold delay is $\tau$ = 44 ms (0.88 m), again demonstrating early warning capability. 

\subsection{Reflection into Shadow Region}
\begin{figure}[h]
\centering
{\includegraphics[width=0.42\textwidth]{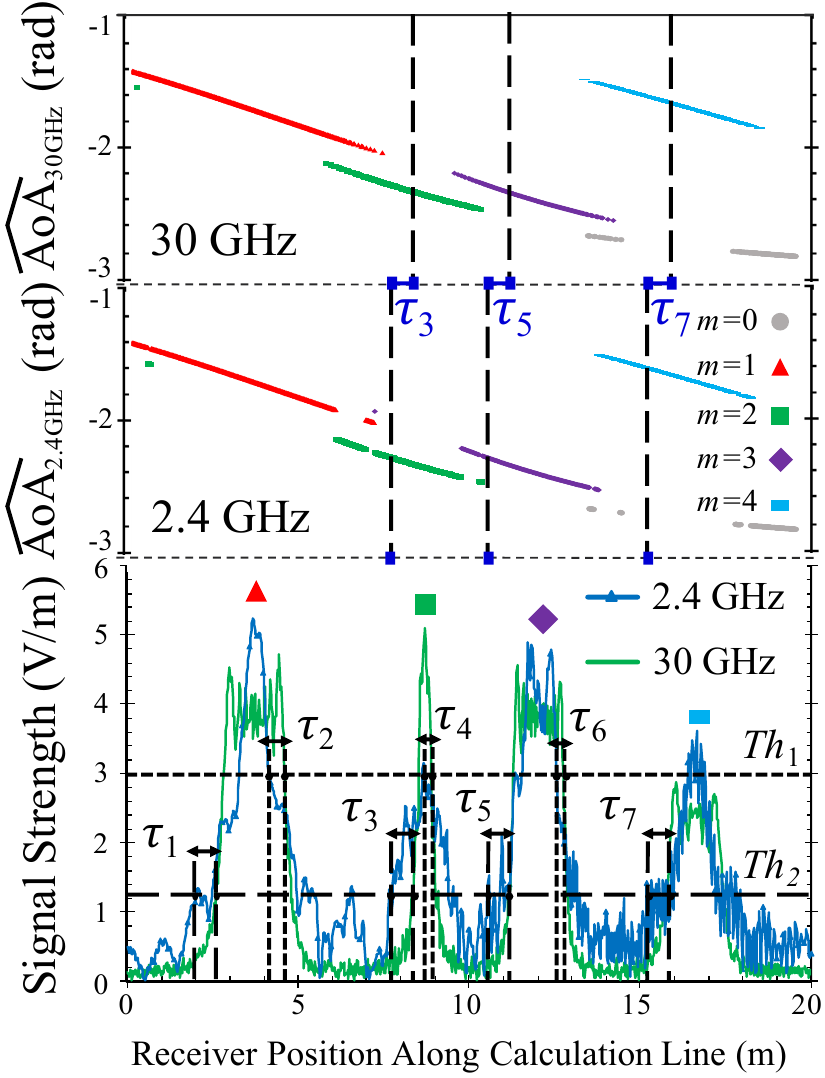}}
\caption{Early warning for strong reflection into a NLoS region and
 $\widehat{AoA}_k(t)$ = $AoA_{\hat{m}}(t)$ where $\hat{m}$ = arg$\max\limits_{m} |h_{k,m}(t)|$ in (\ref{eqnh}). \textbf{Bottom}: Signal strengths were calculated along a 20 m path with reflectors situated and numbered ($m$-value) as in Fig. \ref{LoStoNLoSDiagram}(a). For $Th_1$=70\% and $Th_2$=30\%, $\tau_1(Th_2)$=28.5 ms, $\tau_2(Th_1)$=22 ms, $\tau_3(Th_2)$=31 ms, $\tau_4(Th_1)$=8.57 ms, $\tau_5(Th_2)$=31.75 ms, $\tau_6(Th_1)$=10 ms, and $\tau_7(Th_2)$=21.6 ms (4th mmWave peak does not cross $Th_1$). \textbf{Upper Panels}: $\widehat{AoA}_{f_k}$ along the same path for $f_k$=30 GHz and 2.4 GHz. The vertical dashed lines mark $t_{2.4GHz}(Th_2)$ (middle panel), $t_{30GHz}(Th_2)$ (top panel), and $\tau(Th_2)$ in (\ref{tauth}) for the $Th_2$-crossing of the upcoming strong $|h_{k,m}(t)|$. }
\centering
\label{AllAngleData}
\end{figure}

A strong reflected mmWave signal can enable reliable communications in a NLoS environment, thus alleviating the effects of blockage for mmWave transmission. We explore the early warning of an upcoming strong reflection in NLoS using sub-6 GHz observations in the second scenario, which models a signal being reflected into the region shadowed from LoS as shown in Fig. \ref{LoStoNLoSDiagram}(a). The transmitter aperture has been shortened from 10 m to 5 m to ensure that the calculation path is entirely within the NLoS region. The mobile moves from the left along the horizontal dashed line within the NLoS region and receives a signal from four reflectors at different locations on the mobile path. From left to right, the reflector locations (x, y) in meters are (4.5, -1), (2.5, 1), (6, 2), (18, -3); their sizes are 1.5 m, 0.5 m, 2.0 m, 1.5 m; and their angles counter-clockwise (CCW) from positive x-axis are 135\degree, 120\degree, 120\degree, and 135\degree, respectively. These could represent, e.g, a sign, mailbox, parked car, and dumpster. The sum in (\ref{eqnh}) has five components ($m=0, ..., 4$), and each reflected component $h_{k,m>0}$(t) is sufficiently strong and persistent to produce a viable communication signal for 0.5-2 m or 25-100 ms at vehicle speeds as the mobile crosses the corresponding reflection lobe while the direct component $h_{k,0}(t)$ is weak (NLoS). Early warnings of an upcoming strong reflection ($Th$ = 30\% of average LoS value) and sharp decline of its RSS ($Th$ = 70\%) are investigated, and the delays are shown in Fig. \ref{AllAngleData}(a) and its caption. 

\begin{figure}[!t]
\centering
{\includegraphics[width=0.41\textwidth]{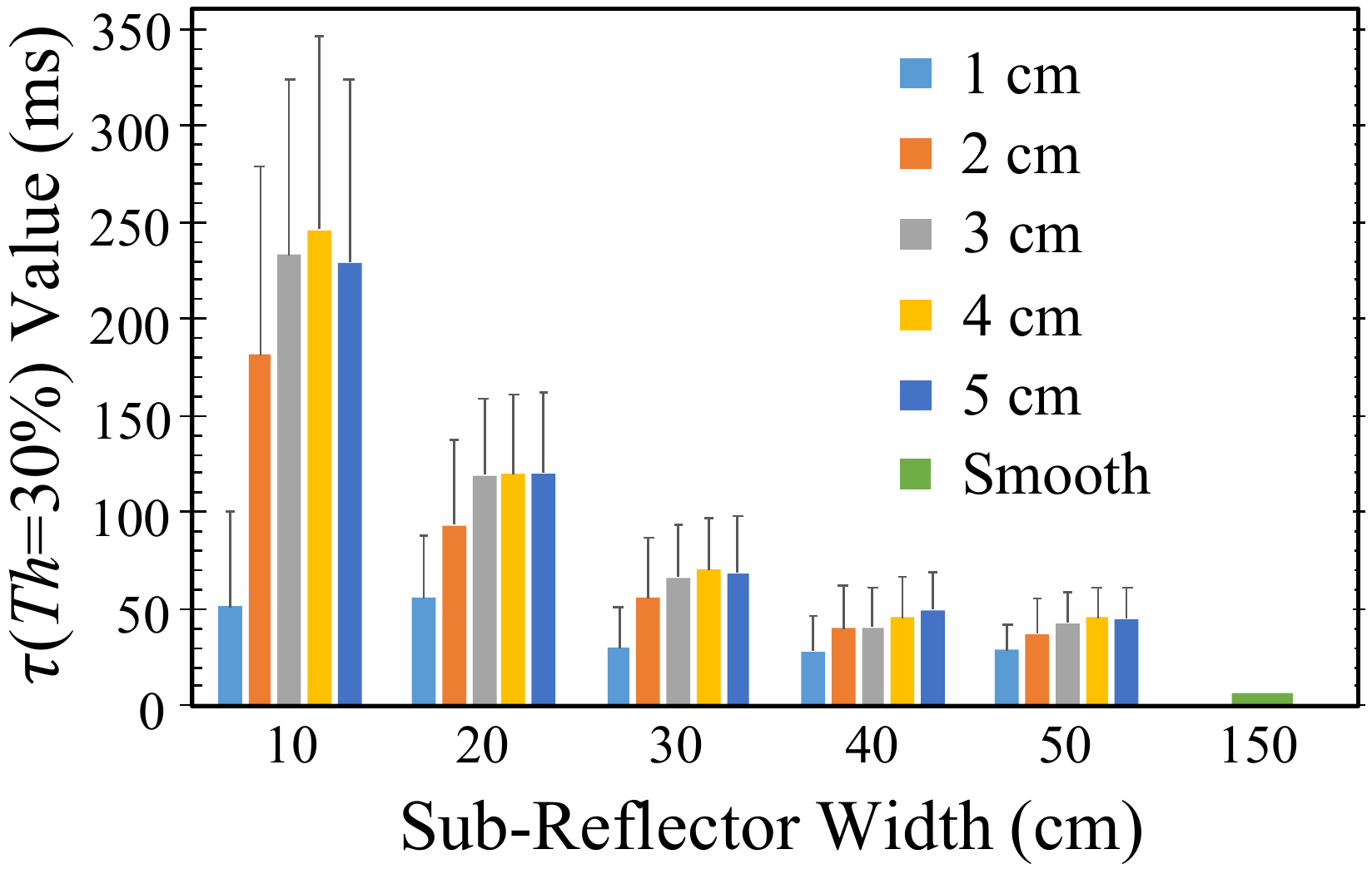}
\hfil
\caption{Effects of a variable rough reflector $m=1$ in the Fig. \ref{LoStoNLoSDiagram}(a) scenario, but with other reflectors removed. The threshold delay $\tau(Th)$ at 30\% of LoS, as the mobile enters the region of that reflection, is shown. Values in each group correspond to sub-reflector perpendicular offset ranges. Different groups correspond to different sub-reflector widths. For comparison, we show at the right the threshold delay (6 ms) of the stand-alone, full-size, smooth reflector.}
\label{RoughReflectors}}
\end{figure}

The shapes of the reflections vs. position in Fig. \ref{AllAngleData} (bottom) vary with the regime of diffraction. \emph{Near-field} diffraction, exhibiting oscillations at the top and a sharp, monotonic decline (in the absence of multipath fading), occurs for $w > \sim 3$ in (\ref{eqnw}), i.e., for a large reflector ($(\xi-\xi_0)$ term), small distances from the reflector ($\rho_m$), or a higher frequency ($f_k$). The diffraction regime can change from near- to far-field by varying any one of these three parameters [SM \S VII]. \emph{Far-field} diffraction is characterized by a broad, featureless lobe, when $w < \sim 0.8$.  The intermediate \emph{transition regime} is marked by large sidelobes. We find that $\tau(Th=30\%)$ is long when the mobile is in the far-field of the reflector at the lower frequency, e.g., the $2^{nd}$ reflection in Fig. \ref{AllAngleData} (bottom), $w_{2.4GHz}$ = 0.8, or when the $Th$ intercepts a sidelobe in the transition regime as for the $1^{st}$ reflector, $w_{2.4GHz}$ = 2.3. When reflectors 2-4 are removed, multipath fading is reduced, and the sidelobe of the first reflection at 2.4 GHz falls below the threshold. The resulting threshold delay (shown in the right-most bar of Fig. \ref{RoughReflectors}) is smaller, $\tau$ = 6 ms, and is due to the wider main lobe at sub-6 GHz compared to mmWave frequencies (as discussed at the end of section II and shown in Fig. \ref{LoStoNLoSDiagram}(c,d)). The main-lobe-induced delay also occurs when both frequencies exhibit near-field diffraction, as for the $3^{rd}$ reflection in Fig. \ref{AllAngleData} (bottom) $w_{2.4GHz}$ = 3.2, $w_{30GHz}$ = 11. The effects of multipath fading are also discussed in [SM \S VIII]. 

\subsection{Angle of Arrival}
In a mmWave beam search, the AoA of the strongest multipath component needs to be estimated. This problem is especially challenging in NLoS environments. Thus, for the scenarios of section III.B, we show the AoA of the strongest component in (\ref{eqnh}), denoted as $\widehat{AoA}_k(t)$ and defined in the Fig. \ref{AllAngleData} caption. We find that $\widehat{AoA}_k(t)$ values change slowly when one multipath component is dominant, due to mobile motion. Note that the $\widehat{AoA}_k(t)$ profiles exhibit a strong degree of congruency for the two frequencies (89.4\%), supporting the findings in \cite{OutOfBand, EstimatingChannels}. For a low-to-moderate density of reflectors, as in Fig. \ref{LoStoNLoSDiagram}(a), the strong reflections are separated by regions of very low RSS, where the mmWave signals are always weaker than the sub-6 GHz signals, due to poorer diffraction. Noise, multipath fading and sidelobes cause the $\widehat{AoA}_k(t)$ instabilities in these low RSS regions. Thus, a practical method to identify $\widehat{AoA}_k(t)$ would employ a RSS threshold or another detection technique. We show the $\tau(Th=30\% \text{ of LoS})$ in all panels of Fig. \ref{AllAngleData}. Note that the $\widehat{AoA}_k(t)$ stabilize after the RSS values cross $Th_2$, and the threshold crossing at sub-6 GHz provides an early-warning for the switch of the $\widehat{AoA}_k(t)$ from one $m$-value to another at mmWave frequencies. 

\subsection{Rough Reflections}
Finally, to provide robustness to the results of section III.B and to simulate realistic building materials and structures, we removed reflectors 2, 3, and 4, and converted reflector $m=1$ in Fig. \ref{LoStoNLoSDiagram}(a) to a rough reflector at the same location by splitting it into many sub-reflectors. Figure \ref{RoughReflectors} shows the results of a statistical analysis as the mobile enters (30\% of LoS) the reflected signal region ($x\sim 1-6$ m) with randomized reflector roughness. Each group of bars corresponds to a different size and number of sub-reflectors, keeping the overall 1.5 m reflector size, and each bar in a group corresponds to a different range for uniform random displacement of the sub-reflectors perpendicular to the surface. The number of components in (\ref{eqnh}) increases with roughness, where roughness is greatest for the group on the left. Within a group, the greatest roughness is for the bar on the right. The mean delay $\tau$  from (\ref{tauth}) and its standard deviation were calculated for 100 iterations of each data point. The statistical analysis in Fig. \ref{RoughReflectors} shows that \emph{roughness increases the early-warning potential} of the lower frequencies significantly. Here, roughness effectively stretches the lower-frequency signal in space while the mmWave signal remains largely unchanged [SM Fig. S3(b)], which is due to the two frequencies experiencing different diffraction regimes. The reduced size of the rough sub-reflectors places the mobile into their far-field regime, with broad diffraction lobes, for the sub-6 GHz signals, $w=0.8, 0.6, 0.5, 0.3, 0.2$ for sub-reflector sizes $(\xi-\xi_0)=0.5, 0.4, 0.3, 0.2, 0.1$ m, respectively, in (\ref{eqnw}), while the mmWave remains mostly in the transition regime due to its larger frequency. In practice, reflectors usually have some degree of roughness, which enhances the early warning potential of the lower frequencies. 


\section{Conclusion}
A Fresnel-diffraction based channel simulation tool was employed to demonstrate that sub-6 GHz signals reach a given threshold much earlier than mmWave signals for a moving receiver. Thus, sudden changes in mmWave signal strength can be predicted using sub-6 GHz observations from several to tens of ms (several to hundreds of slots) ahead in mobile communications systems. Common scenarios, including moving from a LoS region to a NLoS region (blockage), encountering a strongly reflected signal in NLoS, and rough reflections were employed to demonstrate this predictive capability. Future work will focus on measurement-based verification and implementation of robust prediction schemes and adaptive transmission, beamforming, and anti-blockage methods for hybrid 5G systems based on the early warning capabilities of the sub-6 GHz signals.

\ifCLASSOPTIONcaptionsoff
  \newpage
\fi

%
\end{document}


\title{Supplementary Material for "Early Warning of mmWave Signal Blockage and AoA Transition Using sub-6 GHz Observations"}

\author{Ziad~Ali,~\IEEEmembership{Student Member,~IEEE,}
        Alexandra Duel-Hallen,~\IEEEmembership{Fellow,~IEEE,}
        and~Hans~Hallen,~\IEEEmembership{Fellow,~APS}
\thanks{Z. Ali and A. Duel-Hallen are with the Department of Electrical and Computer Engineering, North Carolina State University, Raleigh,
NC, 27607.}
\thanks{H. Hallen is with the Department of Physics, North Carolina State University, Raleigh, NC, 27607.}}

\markboth{IEEE Communications Letters~2019,~DOI:~10.1109/LCOMM.2019.2952602}{}

\maketitle


\IEEEpeerreviewmaketitle

\section{Introduction}
\IEEEPARstart{T}{his} material is organized as follows. In Section S.II, we describe the factors relevant in the choice of a threshold level of RSS (received signal strength) change in the sub-6 GHz band for predicting the mmWave signal change. We show here what was mentioned in the paper: that the 50\% LoS signal value cannot be used.  Section S.III contains the details of the double-diffraction method used for the physical model, with the source of a few of its insights. In Section S.IV, we derive the 'angle of diffraction' from two perspectives. It gives a useful qualitative viewpoint on diffraction and its frequency dependence.  Section S.V contains statistics and additional details of reflection into the NLoS region. The reflector-parameter dependencies of the prediction range are detailed in Section S.VI. In Section S.VII, additional details on the diffraction from a 'small' reflector are provided. In particular, we use plots and model insights to show how changes in wavelength and distance from the object can induce similar effects. Finally, in Section S.VIII, we present examples of topologies with significant multipath fading effects and identify cases where an early warning method needs to be supplemented with multipath fading prediction. Throughout the supplementary material, references to sections, equations and figures within supplementary material will be preceded by an 'S,' as in Fig. S.3(a), while references to such objects in the main paper will not have any special notation, as (6) or Section III.B.

\section{Threshold Levels}
In Section III of the paper, we employ the RSS threshold to predict (provide early warning of) upcoming signal blockage or recovery. There is no optimal choice for the threshold level except in particular scenarios. This is evident in Fig. 2-3 in the paper, and will be more thoroughly justified in Section S.VII. The issues arise from oscillations induced by the diffraction process: (1) the presence of oscillations occurs at low signal level when the mobile approaches the region of a reflection when it is in the transition regime (see Section III.B in the paper); and (2) oscillations occur at large signal levels near the edges of the LoS region or when in the near field of a reflector; or oscillations induced by (3) multipath fading [SM \S S.VIII]. Oscillations due to mechanism (1) can aid the early warning threshold technique. The other types can complicate the process. The threshold should not be too close to the maximum RSS due to uncertainties in its estimation and due to noise. Noise also limits the minimum signal strength that can be used for a threshold. As we show below, the midpoint (50\% LoS) is also inappropriate. In the paper, we employ a threshold signal level that is a 30\% deviation from the nominal value (e.g., below the average LoS signal level or above the average NLoS signal level, the latter being $\sim$0).

Next, we discuss how to determine the average LoS value. The challenges include: (1) obtaining a sufficiently long observation of the LoS signal; (2) identifying suitable observations for setting the threshold value, given that the LoS signal changes with distance from the transmitter as $1/r^2$ and the possibility of partial transmitter blockage as discussed in Section S.IV below. Note that very far from the transmitter, small changes in distance do not impart a very large change in the $1/r^2$ free-space signal drop-off. In this case, $|h_k(t)|^2$ at the NLoS to LoS transition will oscillate about an average LoS value, resembling the oscillations of the strong specular reflection of the mmWave part of Fig. \ref{AllSignalStrengthDat}(b) from 5.5-6.5 m. Here, the average LoS value is computed by averaging the signal during such edge oscillations. When the transmitter is closer, the $1/r^2$ drop-off is relatively important, as the right sides of Figs. 2, \ref{AllSignalStrengthDat}(b, c) at both frequencies, where the transmitter is only $\sim$20 m from the mobile at the transition, and the mobile path length is also 20 m. This proximity of the transmitter would not be expected for sub-6 GHz systems, but is not unexpected for the microcells anticipated for mmWave systems. We chose a nearby transmitter location in this simulation to point out how microcells differ from sub-6 GHz cells.

Finally, we explain why the 50\% amplitude point of edge diffraction is frequency-independent. The position at which the signal reaches 50\% of the LoS RSS for single-edge diffraction occurs when the Fresnel integral argument $w$ in (4) is zero. Since frequency enters this argument as a multiplicative (inverse square root) factor, it drops out when the argument is zero. Thus, 50\% LoS RSS is independent of frequency, so should not be used as a threshold. This is observed in Fig. 2 and later figures as a crossing of $h_1(t)$ and $h_2(t)$ when they reach the 50\% LoS value. Some of the small-reflector data in Fig. 3 does not share this property. Both sides of the reflector are contributing to the diffracted signal, so that the single-edge diffraction is no longer applicable. We seek a prediction threshold that will apply in all cases, so must avoid the 50\% LoS value.

\begin{figure}[!t]
\centering
a.{\includegraphics[width=0.21\textwidth]{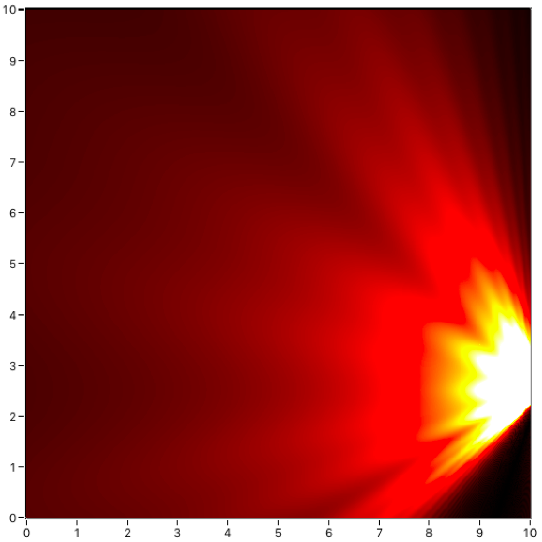}}
{\includegraphics[scale = 0.2]{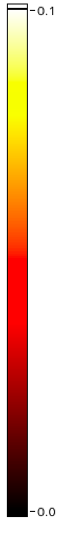}}
b.{\includegraphics[width=0.21\textwidth]{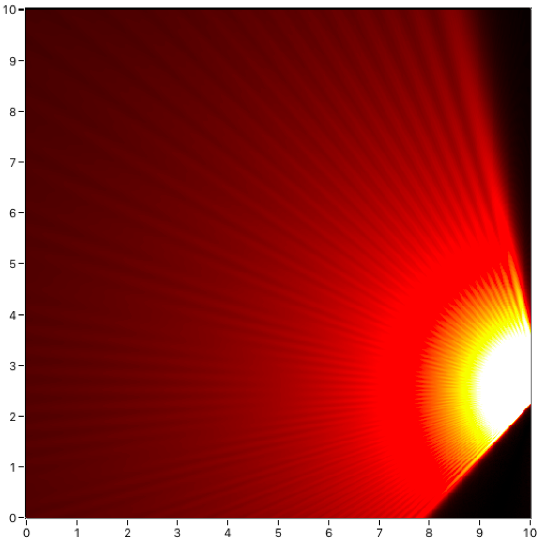}}
\hfil
\caption{The weaker, spreading signals from curved reflectors are demonstrated by taking the same set-up as Fig. 1(c, d) in the paper, but making the reflectors curved with a 0.25 m radius rather than flat. The signal strengths away from the reflector are so much smaller that the range of the color scale here is only 1/10th that used in the paper Fig. 1(c,d). (a) An example of the spatial pattern, $|h_{k,m}(x. y)|$, in a  10 m x 10 m region with a 2.4 GHz transmitter. (b) The same scenario as (a), but $f_k$ = 30 GHz.}
\label{SingleReflect}
\end{figure}

\section{Diffraction Calculation Details}
Two diffractions are required for each reflector in the calculation of $h_{k,m}(t) = h_{k,m}(x(t), y(t)) = h_{k,m}(x_0+v_xt, y_0+v_yt)$ in (1) via the receiver's starting ($t=0$) position $(x_0,y_0)$ and speed $\textbf{v} = (v_x,v_y)$. The first is through an aperture of the transmitter to the center of each reflector, and the second is from the effective source (explained below) behind the reflector to the calculation point. The first diffraction is required to model the LoS to NLoS transition as discussed and seen in Section III.A and Figs. 1(a) and 2 of the paper. The physically common region in which the LOS has been blocked but other signals have been reflected or diffracted (in the calculation region) is investigated in III.B,D and Fig. 3 of the paper. In the case of a large reflector spanning the LoS/NLoS boundary, or close to it, the calculation of the transmitter diffracting to the center of the aperture will fail. Our model relies on the user breaking that reflector into several smaller, adjacent parts (with diffraction calculated to the center of each), which will then provide an adequate modeling of the scenario. A reflection to the point of calculation involves the second diffraction. To simplify this, we place an effective source behind the reflector and let the signal diffract through the aperture, which is the front surface region of the reflector. The field pattern is the same as reflecting the transmitter, including a strong specular component if the reflector is large enough. Diffraction is modeled with Fresnel methods in all cases. The effective source of a flat object, as guided by the method of images, is on the line through the transmitter and perpendicular to the plane containing the reflector front face, the same distance behind that plane as the transmitter is in front of it. Curved objects are accounted for by positioning the effective source closer to the reflector position, along the line containing the transmitter and reflector positions. For the paraxial case, this should be a distance $R/2$ behind the surface, with $R$ the radius of the reflector. In practice, the paraxial approximation is not usually valid, and the effective source should be much closer to the surface. A good compromise is to use $R/4$, which we do. 

By using diffraction in all cases, we do not need to separately treat the LoS and NLoS cases, as they accurately transform into each other with the same equation. Likewise, any size reflector is accurately modeled and none are missed. We state in Section II, paragraph 2 of the paper that curved reflectors are relatively unimportant compared to flat reflectors. We show this in Fig. \ref{SingleReflect}, where we start with the same simulation set-up as for the flat reflector case in Fig. 1(c, d), but make the 0.5 m reflector have a 0.25 m radius of curvature rather than being flat. The transmitter is still far to the left, and the reflector is still in the same position as there. No strong specular central lobe is visible at either frequency. Instead, the signal is strong only adjacent to the reflector, and spreads while decreasing in strength in most directions. The qualitative reason for this is that curved reflectors spread the signals over a large area (think of reflection from a curved mirror of + or - curvature a few radii away). The power is spread over a large area so the power per area is greatly reduced. A few diffraction effects are observed due to the small size of the reflector. The much reduced magnitude of the signal from the curved reflectors is evident in the range of the color scale being only 0.1 rather than 1 for both frequencies. 

We state in Section II, paragraph 2 of the paper that only a 2-d calculation is needed to obtain the qualitative behavior of the sub-6 GHz vs. mmWave signals. In particular, we state that "3-d is more complex but unlikely to result in qualitative differences" in Section II, paragraph 2. We note that the 3-d matters quantitatively, but the point of this paper is to demonstrate the early-warning capabilities of the sub-6 GHz signals. For that problem, the direction from which the reflection comes or the orientation of the blocking surface is not of qualitative importance. The ground bounce, which appears in 3-d scenarios, becomes important in multipath fading, especially in LoS due to the well-known destructive interference (multipath fading) of the LoS and ground-bounce components. However, as we stated in the paper at the end of the 2nd paragraph of Section II : "Scenarios with a LoS component along the entire path are not considered since the LoS component dominates, making these paths useful but not interesting." 

The equations used to calculate the electric field amplitude are given in Section II of the paper. To calculate the received and transmitted powers at the antennas, we must use the Poynting vector $S=c\epsilon_0E^2/2$. The received power is written in terms of the Poynting vector and the antenna effective area $a_{eff} =a_0G_{rx}$ for $a_0$ the area of a perfectly isotropic antenna and $G_{rx} = g_{rx}^2$ the antenna gain in the direction towards the reflector, as $P_{rx}=S a_{eff}=S a_0G_{rx}=S\lambda_k^2G_{rx}/(4\pi) = c\epsilon_0\lambda_k^2\left|{\sum_m{g_{rx,m}E_{k,m}}}\right|^2/(8\pi)$. Note the $m$ subscript on $g$ to denote reflector-dependence, since the direction of arrival at the antenna (hence gain) depends upon the particular reflector. Similarly, we can write the expression for the transmitter Poynting vector in two ways: $S_{tr}=G_{tr,m}P_{in}/(4\pi r_m^2)=c\epsilon_0A_{tr,m}^2/(2r_m^2)$, so the $A_{k,m}(t)$ that is used in (1) can be solved for as $A_{k,m}(t) = \sqrt{G_{tr_m}P_{in}/(2\pi c\epsilon_0)}$. The calculation in (2) assumes a unit input power, with the actual power provided by the $A_{k,m}(t)$ term in (1). The $k$, or frequency, dependence is included since the antenna gains are expected to be frequency dependent. Comparing the received power with the $A_{k,m}(t)$ and (2) \& (3), we find that the output power will depend upon the input power and the product of the antenna gains, scaled by the diffraction and perhaps interference (multipath fading) with other $h$ components with different $m$.

\section{Interpretation of LoS and Angle of Diffraction of an Edge}
We refer to LoS often in the paper, but treat the non-reflected component continuously via diffraction in all regimes. It is therefore useful to give a firm definition of where the signal is LoS. We define the LoS region as that in which the  $h_{k,0}(t)$, the 'direct' or non-reflected component, in (1) is independent of frequency. This will be true if the volume contained in the first Fresnel zone \cite{durgin_practical_2009} for both Rx and Tx antennas is empty, i.e., the receiver is not blocked from the antenna. The outer edge of the first Fresnel zone is the surface for which the path length difference $\Delta r$ between the straight line Tx to Rx distance and the sum of the distances from Tx to the diffraction edge plus the edge to Rx is given by $\lambda_k/2$. The position of the edge is moved parallel to the Tx to Rx line to map these points. The reference shows that $\Delta r = (\lambda_k/4)\,w_u^2$ and that the offsets $u-u_0$ from straight line Tx to Rx to ensure constant $w_u$ (constant Fresnel integral value) at this edge form an ellipse with foci at the transmitter and receiver. The volume within this ellipse should be empty to insure LoS propagation. 

\begin{figure}[!t]
\centering
{\includegraphics[scale = 0.9]{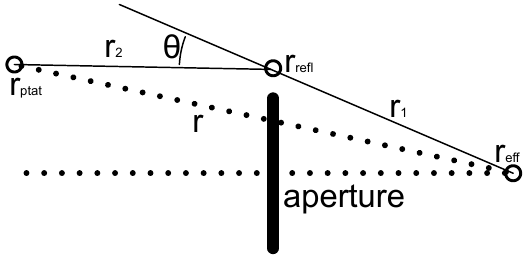}
\hfil
\caption{Schematic drawing to define the diffraction angle parameters.}
\label{diffangle}}
\end{figure}

We also use the term 'angle of diffraction' significantly in our heuristic descriptions of diffraction, such as in the paper at the end of Section II. Consideration of the main lobe of the reflection in Fig. \ref{SingleReflect} strongly suggests that the 'size' of the reflection increases linearly with distance and thus can be characterized by a diffraction angle. This angle is frequency-dependent, increasing for lower frequencies. We now take an analytical approach, with approximation, to derive this result. We begin by noting that the diffraction signal will be constant (up to a factor of $1/r^2$) when the argument $w$ of the Fresnel integral is constant. We thus expect to be approximately on an ellipse as noted previously. There will be a diffraction angle when the measurement point $\mathbf{r}_{rx}$ is shadowed by the edge, as in the schematic Fig. \ref{diffangle}, which defines the quantities we will use. We take the diffraction angle $\theta$ as that between the $\mathbf{r}_{refl}-\mathbf{r}_{eff}$ vector (of length $r_1$), from effective source to the reflector, and the $\mathbf{r}_{rx}-\mathbf{r}_{refl}$ vector (of length $r_2$), from the reflector to the calculation point. Together they form a triangle with the vector $\mathbf{r}_{rx}-\mathbf{r}_{eff}$ of length $r$. We apply the law of cosines to this triangle with the  angle $180^\circ - \theta$ to arrive at
\begin{equation}
r^2 - r_1^2 - r_2^2 = -2r_1r_2cos(180^\circ-\theta) = 2r_1r_2cos(\theta),\label{lawofcosines}
\end{equation}
then expand the following expression:
\begin{equation} \begin{split}
[r+(r_1+r_2)][r - (r_1+r_2)] = r^2-r_1^2-r_2^2 - 2r_1r_2 \\
[2r][\lambda w^2/4] = 2r_1r_2(cos(\theta)-1) = 2r_1r_2 (1-\theta^2/2 - 1). \label{getangle}
\end{split} \end{equation}
On the second line, we used the fact that $r_1+r_2 \sim r$ in the first bracket on the left side, and the results from above that $\Delta r \sim \lambda w^2/4$ for the second bracket, while the law of cosines, (\ref{lawofcosines}), along with a small angle approximation, was used to simplify the right side. We then can solve for the diffraction angle in radians,
\begin{equation}
\theta_{diffraction angle} = w\sqrt{\frac{\lambda r}{2r_1r_2}}. \label{diffangle2}
\end{equation}
We see that the diffraction angle does increase with wavelength, as the square root. Recall that we are tracing a surface in which the Fresnel term is constant, implying that its argument $w$ is a constant here (determined by the particular signal value we set the diffraction angle at).

\begin{figure}[!t]
\centering
a.{\includegraphics[scale = 0.3]{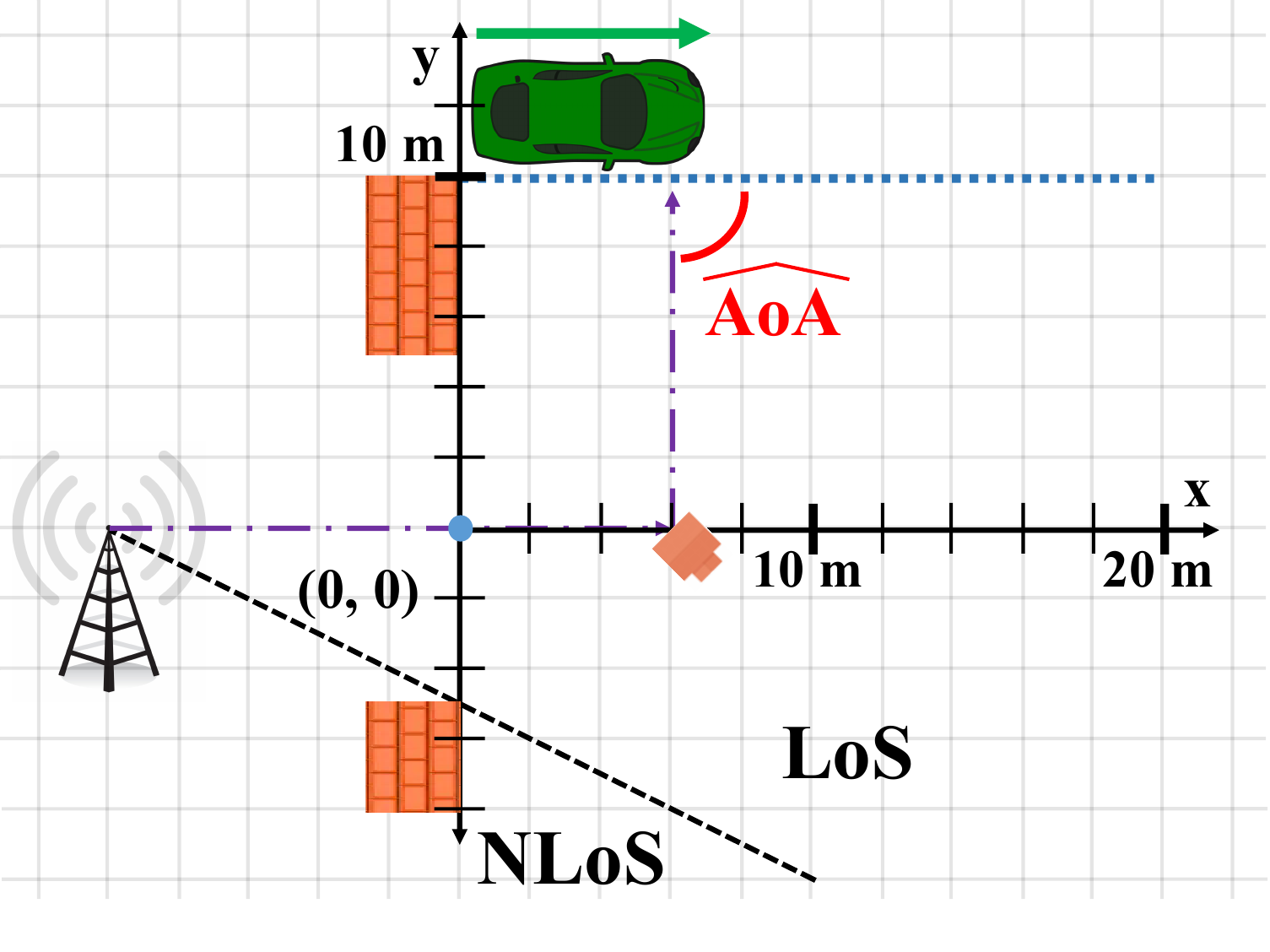}}
b.{\includegraphics[scale = 0.3]{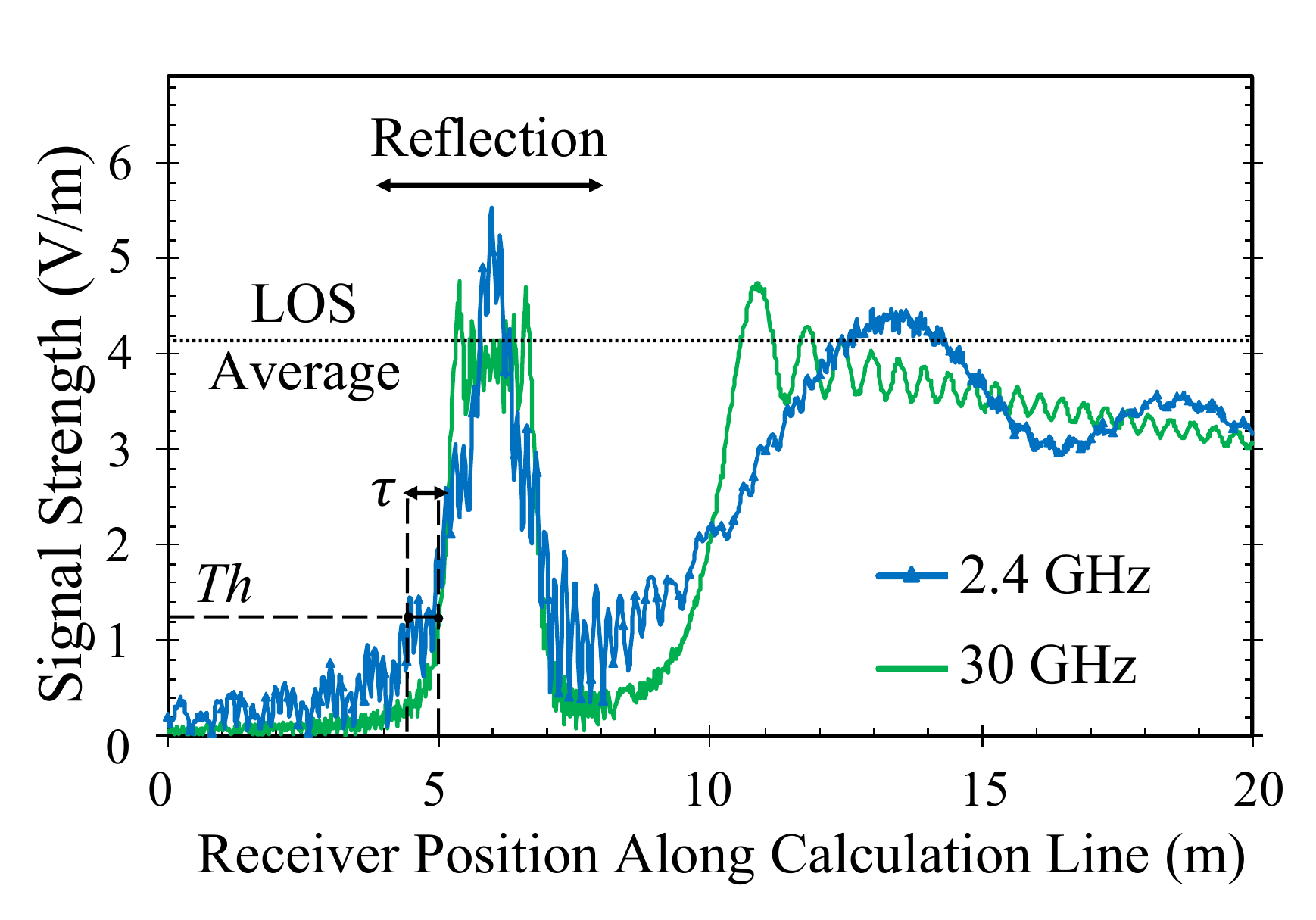}}
c.{\includegraphics[scale = 0.3]{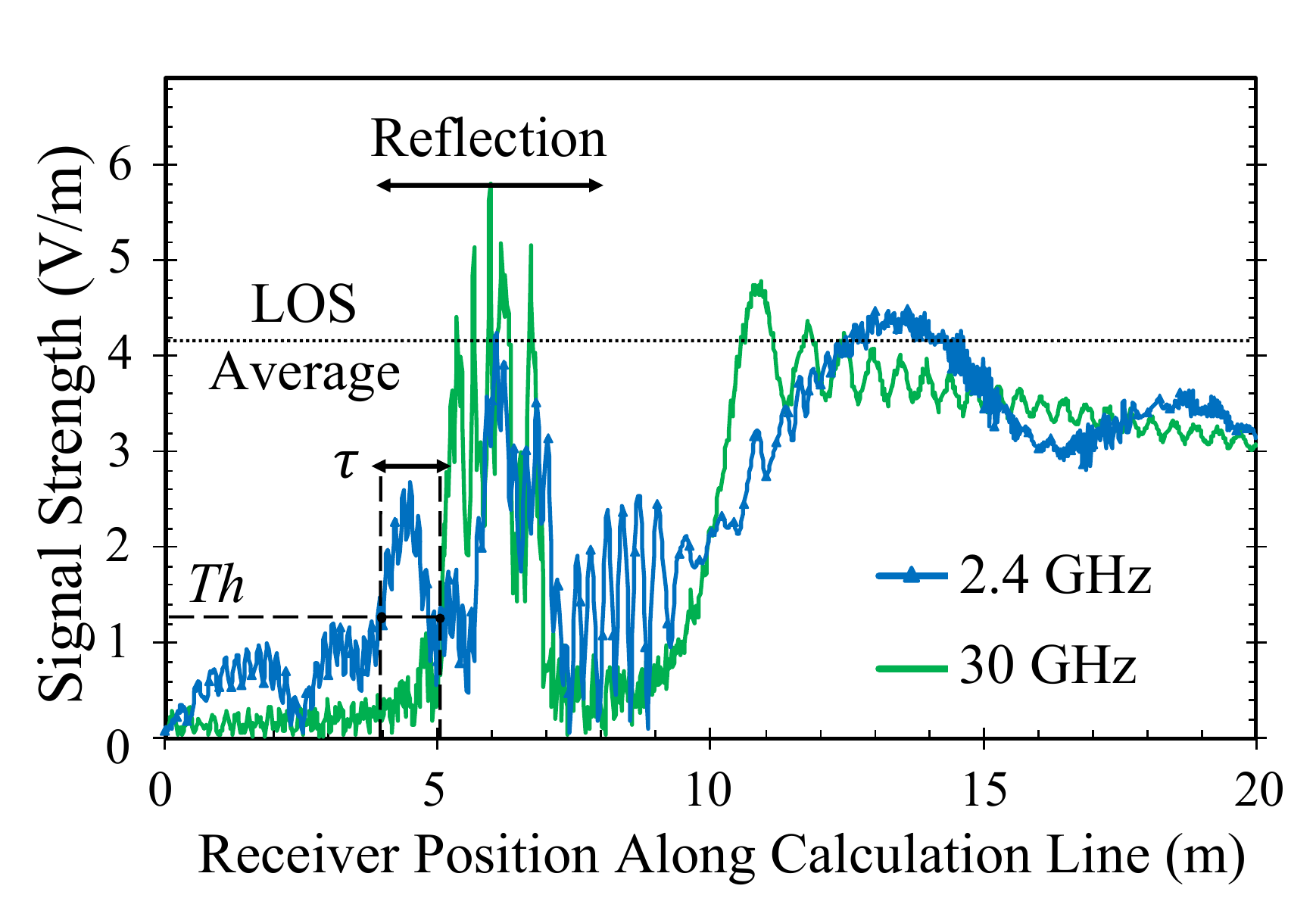}}
\hfil
\caption{RSS $|h_{k}(r)|$ values at $f_1 =$ 2.4 GHz and $f_2 =$ 30 GHz along a 20 m path ($x=$ 0 to 20 m at $y=$ 10 m). (a) Scenario for reflection into a NLoS region; (b) Scenario of (a) for a smooth reflector, $\tau(Th=30\%LoS)$ = 28 ms; and (c) rough reflector, $\tau(Th=30\%LoS)$ = 64 ms.}
\label{AllSignalStrengthDat}
\end{figure}

\section{Details of the Reflection into a NLoS Region.}
We discuss a reflection into a NLoS region in the paper Sections III.B-D. The early warning capabilities of the sub-6 GHz signals for RSS changes in the mmWave signals are demonstrated in a variety of diffraction regimes. Further, the predictive ability of the lower frequencies for the largest mmWave component and its angle of arrival are discussed. Finally, a statistical analysis of the effects of reflector roughness, which helps the early warning abilities, is shown. Here, we show signal amplitude vs. position for two cases near the NLoS/LoS transition, and discuss the effects of reflections into the NLoS on the NLoS/LoS and LoS/NLoS transitions, with a statistical analysis.

The scenario shown in Fig. \ref{AllSignalStrengthDat}(a) models a signal being reflected into the region shadowed from LoS. The mobile moves from the left along the dashed line within the NLoS region and receives a signal from the reflector (size = 1.5 m, 135$\degree$ counter-clockwise angle from positive x-axis) at $x = 6$ m. The sum in (1) reduces to two components ($m=0,1$), and the reflected component $h_{k,1}$(t) is sufficiently strong and persistent to produce a viable communication signal even when the direct component $h_{k,0}(t)$ is weak (NLoS). The threshold delay, $\tau(Th=30\%)$, defined according to (6), is shown in Fig. \ref{AllSignalStrengthDat}(b), and equals 28 ms, implying successful prediction is possible in this case.

To provide robustness to this result and to simulate realistic building materials and structures, we converted the single (smooth) reflector to a rough one by splitting the reflector into many smaller reflectors (size 0.2 m) randomly displaced from each other, perpendicular to the surface, within a range of +/- 0.04 m [SM \S V]. This increased the number of components in (1) from two to eight. As shown in the specific scenario of Fig. \ref{AllSignalStrengthDat}(c), roughness increases the predictive potential of the lower frequencies significantly, by effectively "stretching" the lower frequency RSS in space, causing the distinct sidelobes seen in Fig. \ref{AllSignalStrengthDat}(c) to occur well before any substantial increase in the strength of the mmWave signal. Essentially, the large sidelobes imply that we are in the transition regime of the rough reflector at lower frequencies while the high-frequency signal remains closer to the near-field, low-sidelobe regime. We performed a statistical study on reflector roughness and give the results in Section III.D of the paper.

\begin{figure}[!t]
\centering
{\includegraphics[scale = 0.3]{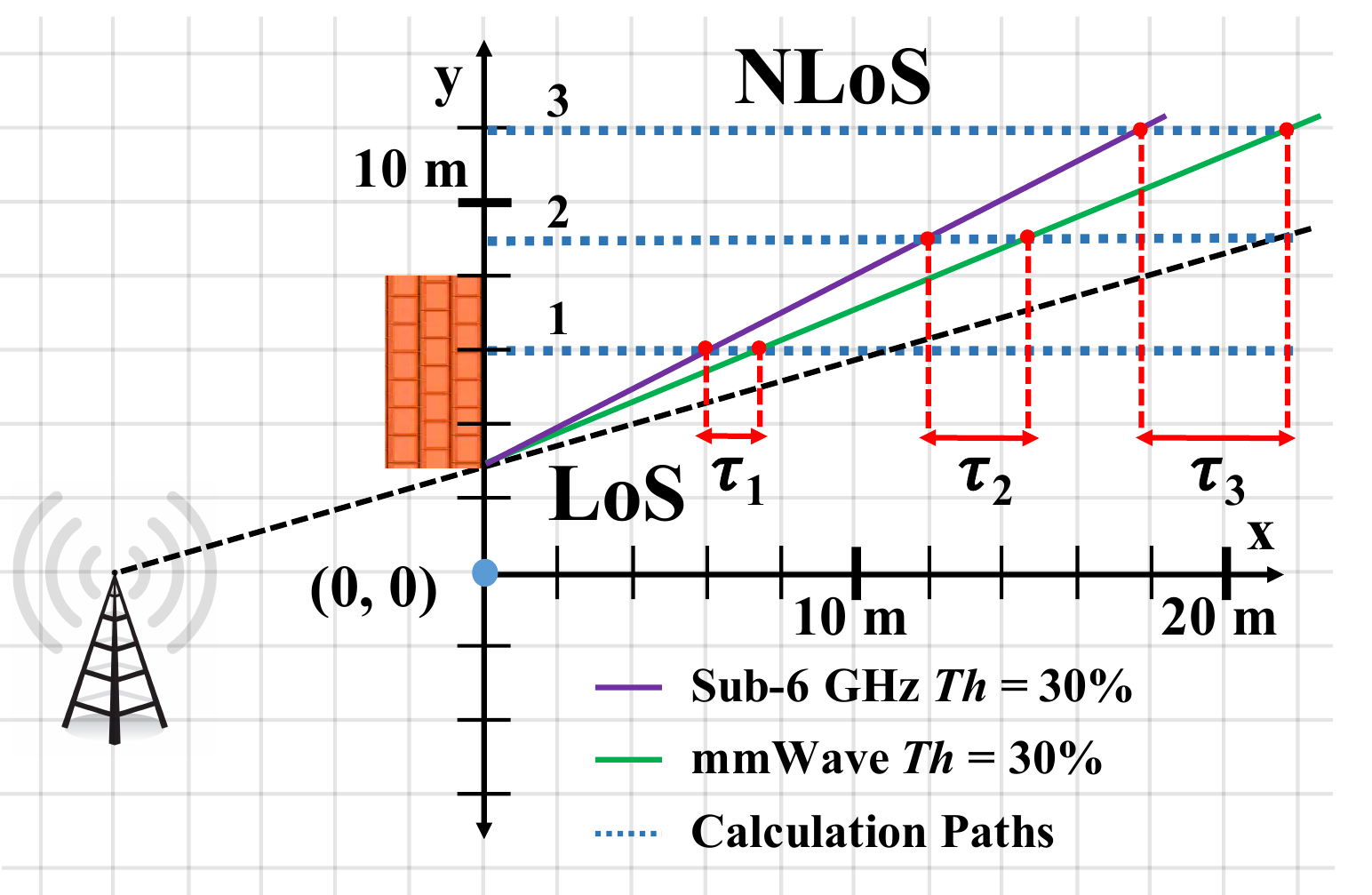}
\hfil
\caption{Depiction of how transmitter distance increases threshold delay between sub-6 GHz and mmWave signals. The purple line represents the approximate locations in space corresponding to the sub-6 GHz signal $Th = 30\%$ threshold, while the green line represents the same for a mmWave signal. The delay ($\tau$) between these lines at the points where they intersect various calculation paths (1, 2, and 3) increases as the distance between the calculation path and the transmitter increases.}
\label{LinesDiagram}}
\end{figure}

\begin{figure}[!t]
\centering
{\includegraphics[scale = 0.32]{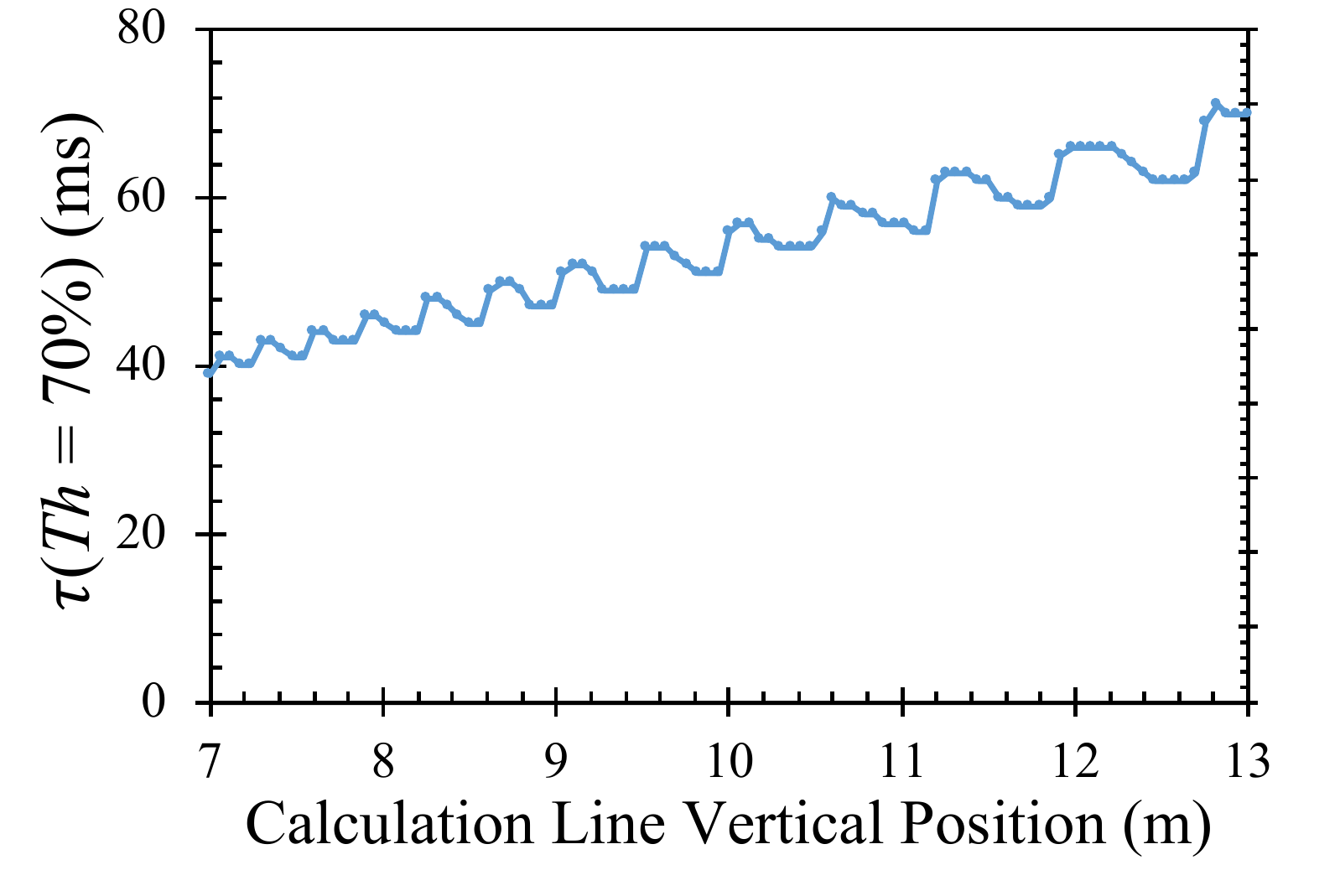}
\hfil
\caption{Dependence of the threshold delay on the distance between the diffracting edge and path. In particular, the y-location of the path (which is along x) is given on the abscissa, see Fig. 1(a). Here, the average LOS value used to calculate the 30\% drop-off is calculated as approximately the maximum LOS value reached for the path at that y-location. This is used since the range over which the actual average value should be calculated changes at every calculation point.}
\label{CalcLineVertical}}
\end{figure}

\section{LoS/NLoS Transition Dependencies on Distance}
In this section, we study the strong, nearly monotonic dependence of the threshold delay (for the NLoS/LoS transition) on the distance between the closest edge of the transmitter aperture and the position on the mobile's path at which the NLoS/LoS transition takes place. Qualitatively, since the diffraction strength stays approximately constant along a particular angle of diffraction, the threshold delay should have an approximately linear dependence on the spacing between the mobile's path and the transmitter aperture edge that is causing the LoS/NLoS transition, as depicted in Fig. \ref{LinesDiagram}. Figure \ref{CalcLineVertical} shows that indeed the threshold delay is nearly monotonic with this spacing, interrupted only by the oscillations of the diffraction curves (see Fig. 2 and \ref{AllSignalStrengthDat}(b, c)). The saw-tooth pattern results from horizontal 'jumps' (in distance) as the threshold crosses the top of an oscillation from that oscillation top to the diffraction curve at the same height to the right. For practical spacing between mobile and the edge, the threshold delay is sufficient for prediction purposes.

\begin{figure*}[!t]
\centering
{\includegraphics[scale = 0.5]{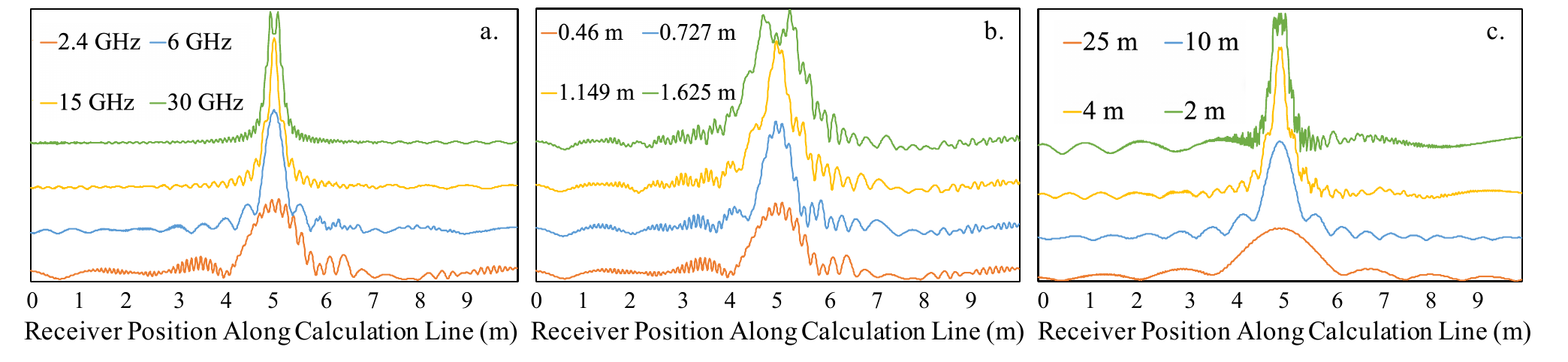}
\hfil
\caption{Parameter variations for a small reflector show the several routes for conversion of the near-field to the far field regime. (a) The wavelength is varied with values 0.125 m, 0.05 m, 0.02 m, and 0.01 m, corresponding to normalized $2(u_2-u_1)^2/(\lambda r_2)$ values of 0.676 (far-field), 1.69, 4.225, and 8.45 (near-field) respectively. (b) The reflector size is varied with values 0.46 m, 0.727 m, 1.149 m, and 1.625 m, having the same normalized values as in part (a). (c) The reflector to calculation distance is varied with values 25 m, 10 m, 4 m, and 2 m, having the same normalized values as in part (a).}
\label{SmallReflector}}
\end{figure*}

\section{Details of Diffraction From a 'Small' Reflector}
In the calculation of (1) or (2), an effective source Fresnel-diffracts through a small aperture for any reflector. The diffraction angle is larger for the lower frequencies (Section S.IV) giving the wider region of increased signal level at lower frequencies. In the paper, we identified three regimes for reflectors in Section III.B: the near-field, the transition region, and the far-field region. Therefore, what we mean by a small reflector is dependent on both the frequency and distance away from the reflector in addition to the actual reflector size. These regions are characterized by: near-field - resembling back-to-back NLoS/LoS transitions with a region in the middle having oscillations about a constant value (a 'flat top') with steep, monotonic edges (unless there is multipath fading present), transition - having large sidelobes and no flat top, and far-field - with broad peaks with smaller sidelobes, see the description and figure references in Section III.B and Fig. \ref{SmallReflector}. We investigate some of these aspects here. Note that multipath fading can insert some oscillations into the response, as we show in [SM \S VIII]. The regime of the reflector impacts the predictive capabilities of the sub-6 GHz signals for the mmWave signals. The sidelobes of the transition region can make prediction difficult, but examples seem to indicate that the sidelobes on the sub-6 GHz signal are larger and trigger the change sufficiently in advance of those of the mmWave, when both are in the transition region. When the sidelobes are not large enough, the main lobe is also wider at lower frequencies, Fig. \ref{SingleReflect}, so early warning is maintained. As for the LoS/NLoS transition discussed in the last section, distance of the mobile's path from the reflector has the largest influence on the threshold delay, as it is one of the three primary instigators of changes in the reflection regime. As evident in (4) and discussed quantitatively below, the other factors are the frequency and the size of the reflector. Larger reflectors have a reflection/diffraction pattern that is large, strong, and has edges similar to the LoS/NLoS transition over a much greater distance from the reflector than the reflection/diffraction pattern of a physically smaller reflector. The smaller the reflector, the closer to that reflector one has to place the path to see a qualitatively similar channel amplitude plot. The latter results from the angle-dependence of two aspects of the reflector: the reflector size (with smaller size producing a larger angle) and the angle of edge diffraction (as a LoS/NLoS transition), such as discussed previously in Section S.IV. The competition results in the conversion of the pattern from near-field through transition to the far-field regime with distance from the reflector. Of course, the larger distance from a reflector means the pattern is much weaker and more spread out, although the sidelobe relative amplitudes and shape of the peak would qualitatively resemble that of a smaller reflector on a closer path. 

To be more quantitative, consider the argument of the Fresnel integral, (4). The dependencies can be characterized in terms of the reflector size ($u_{2,m} - u_{1,m}$), the wavelength $\lambda_k$, and the the distance via $\rho_m = r_{1,m}r_{2,m}/(r_{1,m}+r_{2,m})$ as in Fig. 1(c, d) and in (5), where the reflector ($m$) and wavelength ($k$) dependence has been indicated by the $m$ and $k$ subscripts. The latter term in (5) has limits of the smaller of the two distances when they are greatly different in value, or half that distance when they are equal. Thus, we can consider it to be the smaller distance and be correct to within a factor of 2. Since we have previously discussed (near the end of the paper introduction) that curved reflectors are usually not important, and since flat reflectors have effective source to reflector distance $r_1$ approximately equal to the (relatively large) transmitter-reflector distance, the distance $r_2$ from the reflector to the calculation point will usually be smaller than $r_1$, so $\rho \sim r_{2,m}$, and our approximation is much better than a factor of two. We will look at how the variations in $r_{2,m}$, $\lambda_k (f_k),$ and the reflector size ($u_{2,m} - u_{1,m}$) can combine to create basically the same reflection/diffraction pattern on a path (up to the $1/r^2$ dependence). We do so by creating a series of simulations, shown in Fig. \ref{SmallReflector}, for which each of these parameters is varied from reference values of 5 m, 0.125 m (2.4 GHz), and 0.46 m, respectively. For the simulation in which $r_{2,m}$ is varied, the reference is scaled to $\lambda_k$ of 0.05 m (6 GHz) and reflector size of 0.65 m (same normalized values) to better illustrate the simulated relationship. Both $r_{2,m} \text{ and } \lambda_k$ enter as an inverse square root dependence, so they must be made smaller to engender the change from far field to near field, while $u_{2,m} - u_{1,m}$ enters as a linear dependence, so it must be increased by a lesser fraction to obtain the same result. Since the reference values were chosen to start in the far-field regime, the parameters are changed in this way.

Naively, one might try to normalize the distance $r_2$ in terms of the wavelength or reflector size to predict the regime. We have just shown that this will not work. The three parameters enter together, so one would need to normalize the distance by the square of the reflector size divided by the wavelength. We have done that in Fig. \ref{SmallReflector} (a) to create step sizes in 2*(reflector length)$^2$/(wavelength*distance) that are also taken in parts (b) and (c) of the figure, making the plot similar for the three cases.

Finally, it is useful to consider whether the insights gained above, the relation of the regimes and their characteristic spatial patterns in Fig. \ref{SmallReflector} to the pattern size, can be utilized for prediction. In particular, the relative magnitude of the sidelobes (relative to signal height or to the other frequency's sidelobes) as a reflector enters the signal along a path can be used to estimate the length of time that the reflection will be useful for signaling. This estimate can be made before the signal reaches its full value, thus increasing the value of the prediction.  It also ensures that the mobile accurately knows when the signal from a small reflector will decrease, which is important since the sidelobes interfere with the normal predictive nature of the lower frequencies.  An implementation of an early warning system such as this can also take advantage of various signal changes when a NLoS region approaches while in a LoS region, e.g. the oscillations of the RSS prior to the LoS to NLoS transition, which occur due to the Fresnel integral behavior and have a higher rate of oscillation at higher frequencies. Our model would provide a good testbed for these methods also.

\begin{figure*}[!t]
\centering
a. {\includegraphics[scale = 0.28]{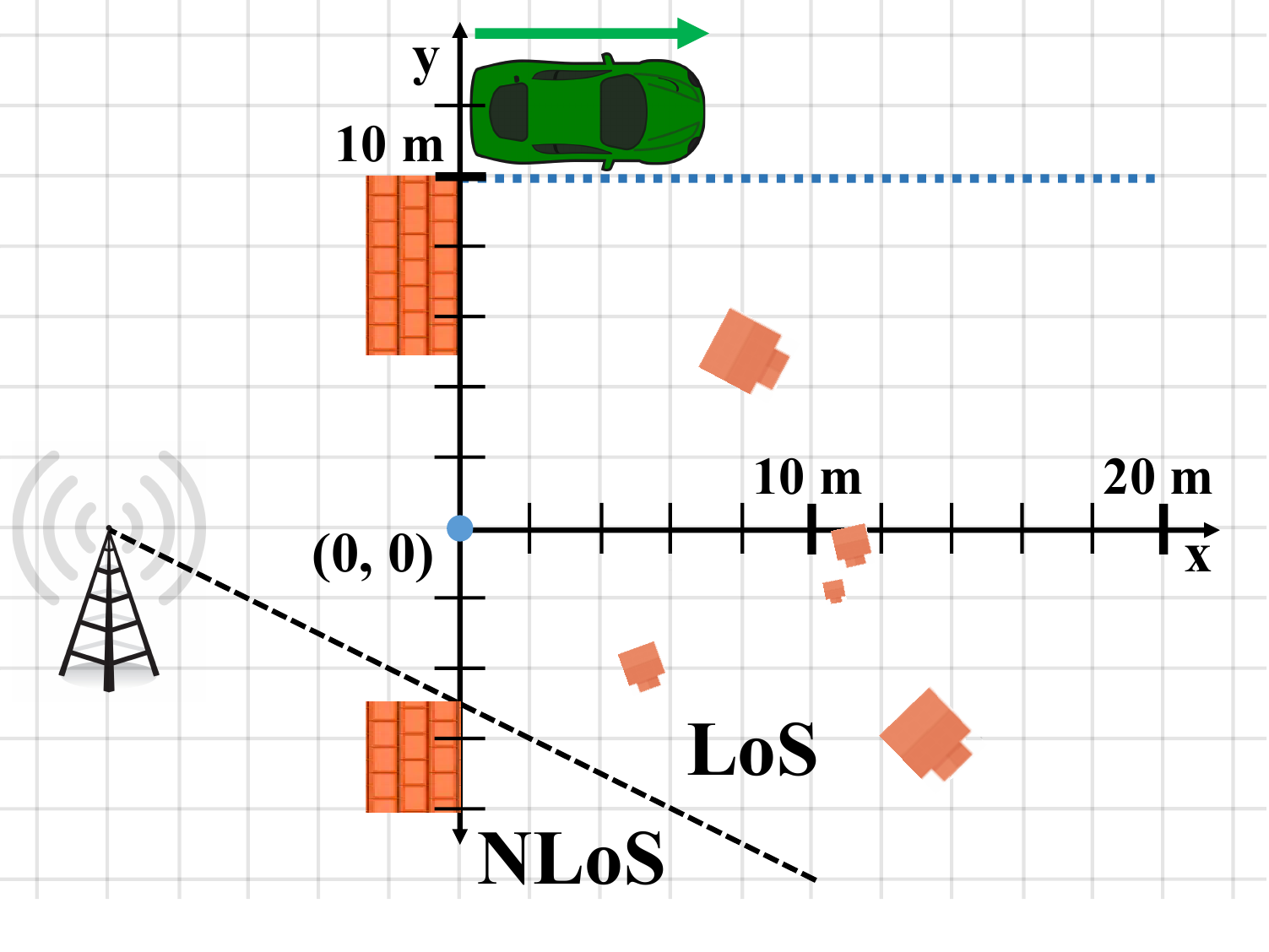}}
b. {\includegraphics[scale = 0.3]{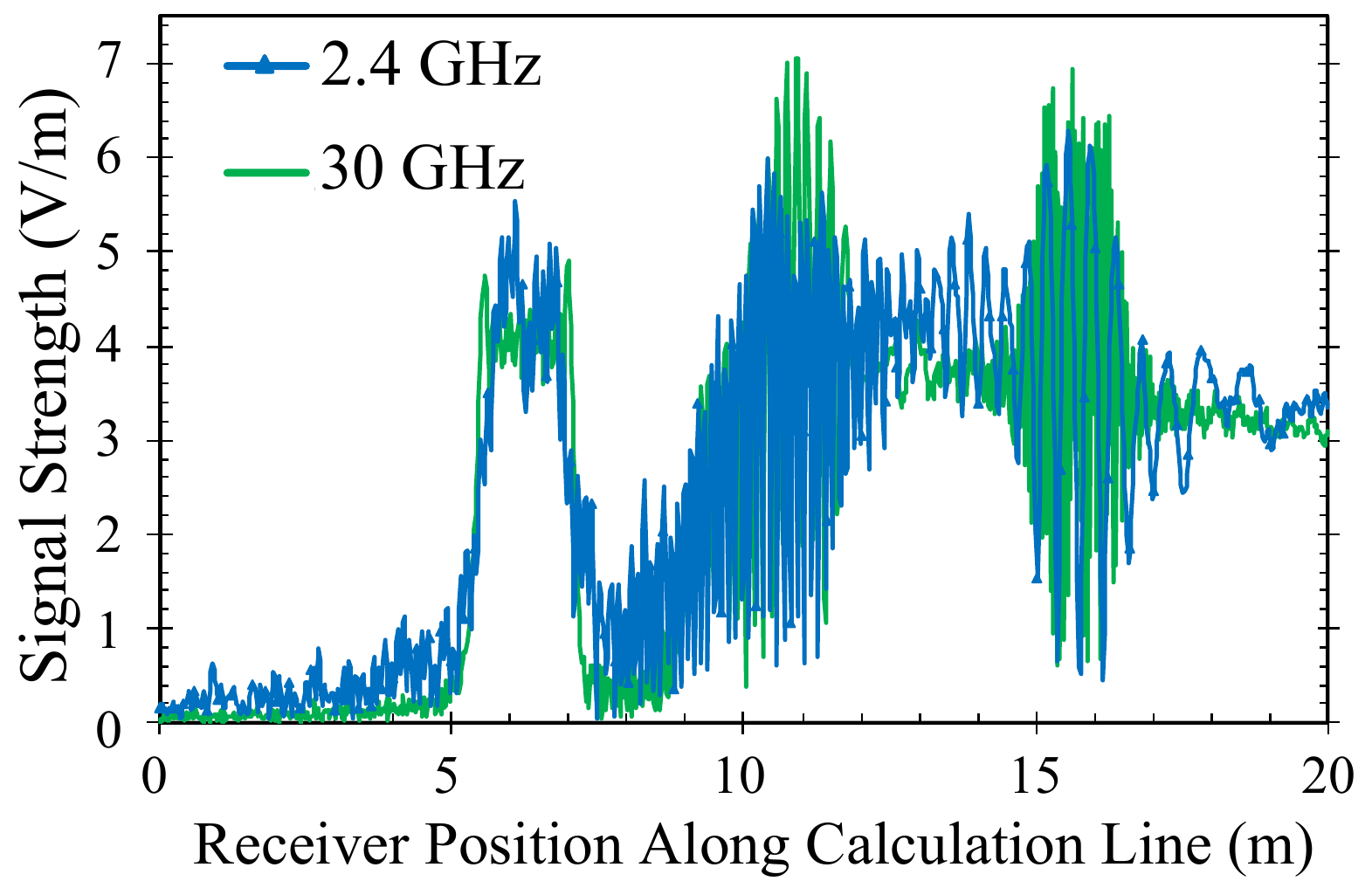}
\hfil
\caption{Reflections into NLoS, the transition, and LoS are shown in one realization of a random reflectors study. (a) The scenario schematic diagram is shown. The path is 20 m long.  (b) The signals along the path at the two frequencies are shown. Three reflections are observed, with two strongly fading. The one on the left is in the NLoS region, so little non-reflected signal is present and fading is minimal.}
\label{flatfading}}
\end{figure*}

\section{Multipath Fading Effects}
Multipath fading is important whenever or wherever there are one or more multipath components in (1) with similar magnitudes. Using the physical model, we identify two interesting cases where multipath fading plays a significant role and discuss approaches to signal prediction for these cases.

\subsection{Reflection into the NLoS/LoS transition region}
In Section III.A, we show the early warning results for blockage and recovery when $M=0$ in (1). When $M>0$, several reflectors are present as in Fig. \ref{flatfading}(a), and multipath components might interfere near the NLoS/LoS boundary. In Fig. \ref{flatfading}(b), one reflection is directed into the NLoS region, similar to those studied in Section III.B of the paper, another reflection is directed fully into the LoS region (at about 16 m), and we observe rapid, frequency-dependent oscillations due to multipath fading effects, with an envelope given by the amplitude of the reflected signal vs. position. The case of interest for this work, since it impacts blockage and recovery, is the reflection that overlaps the NLoS/LoS transition region near 10-12 m in Fig. \ref{flatfading}(b). Again, rapid, frequency-dependent oscillations due to multipath fading effects are observed. The envelope is more complicated since the relative signal amplitudes of the reflected signal vs. the $m=0$ signal vary from one side of the overlap region to the other. The region of stronger fading, when the signals are similar in magnitude, is at a different position for sub-6 GHz frequency than for mmWave frequency, since the mmWave has a narrower NLoS/LoS transition region, as discussed in Section III.A in the paper. This is evident in the color asymmetry for the NLoS/LoS region (10-12 m) fading compared to the symmetry in the LoS region (15-17 m) fading of Fig. \ref{flatfading}(b). As we stated in the last sentence of III.B, long-range fading prediction [12 from paper] can be employed jointly with the proposed early-warning method to reveal the upcoming RSS variations in this case.

\begin{figure*}[!t]
\centering
{\includegraphics[scale = 0.8]{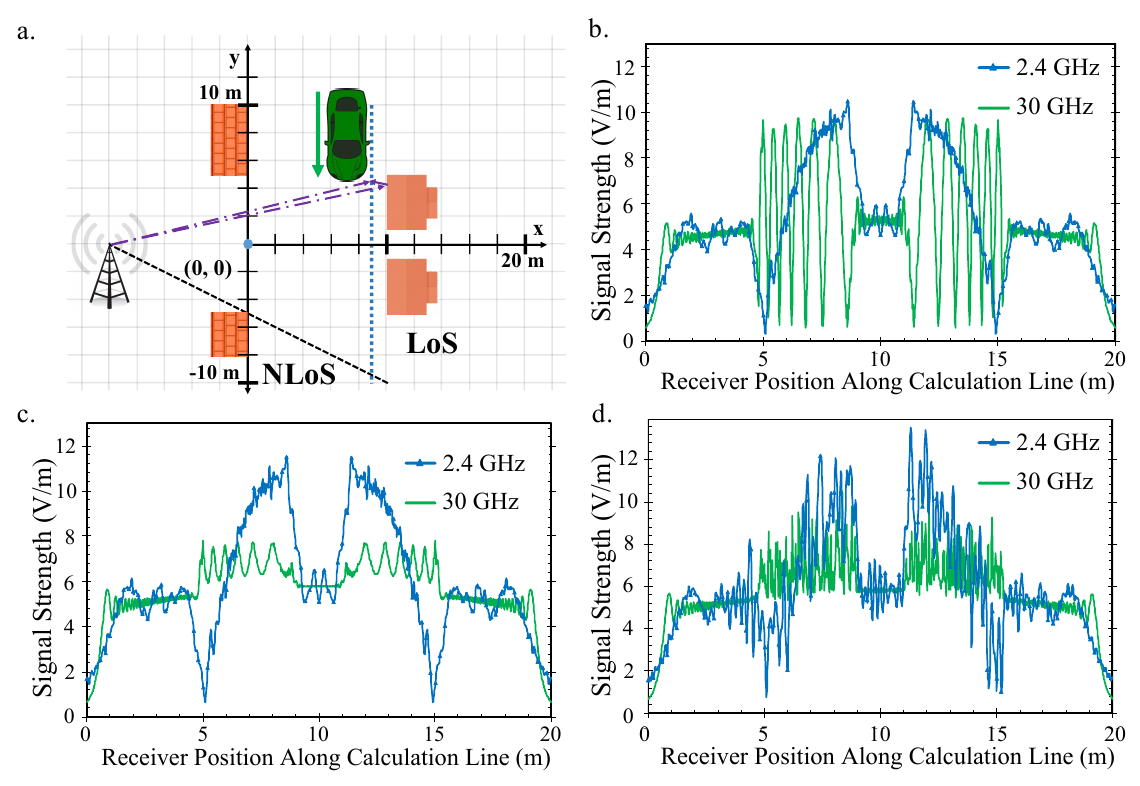}
\hfil
\caption{Multipath of a LoS signal and its back-reflection show unexpected characteristics. (a) The scenario schematic diagram.  (b) Signals along the path when the reflectors are completely flat and the frequencies pure. (c) The effects of introducing a band of frequencies rather than a pure tone. (d) The effects of introducing surface roughness to the reflectors.}
\label{BigFlat}}
\end{figure*}

\subsection{LoS plus near back-reflection from a large flat  object}
We consider an unusual but not irrelevant case of a LOS signal with reflection from a large flat reflector received in almost the opposite direction. Such a case could be expected to be unimportant considering the directionality of mmWave antennas, but antenna arrays do have degeneracies: a 1-d antenna array cannot tell apart AoA on a cone, and a 2-d flat antenna has a left-right degeneracy. Thus, a 1-d antenna with the array length parallel to the reflector surface or a 2-d antenna parallel to the reflector surface would both receive both the signals considered. In this case, the mmWave variations are not anticipated by the low-frequency signals, although the addition of roughness to the reflector or a bandwidth to the signal quench these variations of the mmWave signal and sub-6 GHz predictive behavior is mostly re-established, although there is a possibility of a 'false-positive' indication of signal loss.

Simulated RSS characteristics for 30 GHz (mmWave) and 2.4 GHz (sub-6 GHz) signals were calculated along a receiver calculation path of length 20 m. Two large reflectors are placed perpendicularly to the aperture's normal, with the calculation line placed 1 m in front of them, Fig. \ref{BigFlat}(a). The RSS and reflection characteristics for the 2.4 GHz and 30 GHz signals when both reflectors are completely flat is shown in Fig. \ref{BigFlat}(b). The deep oscillations at the mmWave frequency are due to the slow relative path length variation as the mobile moves in front of the reflector, combined with a reflectivity near unity. Each of the $\sim$6 oscillations corresponds to a relative path length change of one wavelength. The sub-6 GHz signal only shows a partial oscillation (a little less than 1/2 period) since its wavelength is longer than the mmWave by a factor of 12.5, and the relative path length generated over the distance in front of the reflector is not sufficient for an oscillation.  Figure \ref{BigFlat}(c) shows the RSS and reflection characteristics when the pure-tone frequency is replaced by a band (0.4\% of signal frequency). This could reflect the bandwidth of data on the signal, for example. Figure \ref{BigFlat}(d) shows the RSS and reflection characteristics when roughness (0.1 m sub-reflector widths, 0.01 m maximum uniform random sub-reflector offset perpendicular to the reflector surface) is introduced onto the reflectors after breaking them into small sub-reflectors. This figure demonstrates that in this type of scenario, while the 2.4 GHz signal does not serve as an adequate early-warning indicator in the perfect-world problem, introducing real-world factors (roughness, averaging) mitigates most of the destructive reflective effects experienced by the mmWave signal, and results in only a false-positive indicator of mmWave signal loss since the longer wavelength signal at 2.4 GHz is less sensitive to combined bandwidth addition and simulated roughness than the mmWave signal.

A realistic scenario where the results of this section apply is that of a mobile user walking in front of a large store-front window when the base station is directly across the street. The large, flat window provides a strong specular reflection in the mmWave spectral region. The mmWave oscillations are reduced as the reflection becomes less of a back-reflection, that is, as the mobile moves further down the street. Our example shows that there is only minimal reduction in these oscillations after 5 meters of travel (where the region of specular reflection from the modeled reflector ended). The effect described in this section would be important for tens of meters if the window or windows were that long.

Note that the oscillations observed in this example are due to multipath fading (not blockage or recovery) and can be predicted using, e.g., the methods described in [14 from the paper].

\ifCLASSOPTIONcaptionsoff
  \newpage
\fi
\bibliographystyle{IEEEtran}
%

%
